\documentclass[aps,prb,groupedaddress]{revtex4}
\usepackage{graphicx}
\usepackage{color}
\usepackage{epstopdf}
\usepackage{epsfig}

\begin{document}

\title{Measurement of the Current-Phase Relation in Josephson Junctions Rhombi Chains}

\author{I. M. Pop, K. Hasselbach, O. Buisson, W. Guichard and B. Pannetier}

\affiliation{$^1$Institut N\'eel, CNRS, 25 Avenue des Martyrs, 38042 Grenoble Cedex 9,
associated with University Joseph Fourier and Institut National Polytechnique de Grenoble}

\author{I. Protopopov}

\affiliation{$^2$L. D. Landau Institute for Theoretical Physics, Kosygin str. 2, Moscow 119334, Russia}

\date{\today}

\begin{abstract}
We present low temperature transport measurements in one
dimensional Josephson junctions rhombi chains. We have measured the
current phase relation of a chain of 8 rhombi. The junctions are
either in the classical phase regime with the Josephson energy
much larger than the charging energy, $E_{J}\gg E_{C}$, or in the
quantum phase regime where $E_{J}/E_{C}\approx 2$. In the strong
Josephson coupling regime ($E_{J}\gg E_{C} \gg k_{B}T$) we observe
a sawtooth-like supercurrent as a function of the phase difference over the chain. The period of the supercurrent oscillations changes abruptly from one flux quantum $\Phi_{0}$ to half the flux quantum $\Phi_{0}/2$ as the rhombi are tuned in the vicinity of
full frustration. The main observed features can be understood
from the complex energy ground state of the chain. For
$E_{J}/E_{C}\approx 2$ we do observe a dramatic suppression and
rounding of the switching current dependence which we found to be
consistent with the model developed by Matveev et
al.(Phys. Rev. Lett. {\bf 89}, 096802(2002)) for long Josephson junctions chains.
\\
PAS number(s): 74.40+k, 74.50+r, 74.81.Fa, 73.23b
\end{abstract}

\maketitle

\section{Introduction}
Arrays of small Josephson junctions exhibit a variety of quantum
states controlled by lattice geometry and magnetic frustration
\cite{Fazio_2001}. A particularly interesting situation occurs in
systems with highly degenerate classical ground states where non
trivial quantum states have been proposed in the search for
topologically protected qubit states\cite{Ioffe_2002}. The
building block for such a system is a rhombus with 4 Josephson
junctions and the simplest system is the linear chain of rhombi as
proposed by Dou\c{c}ot and Vidal\cite{Doucot_02} along the line of
the so-called Aharonov-Bohm cages \cite{Vidal_98}. The main
consequence of the Aharonov-Bohm cages in the rhombi
array is the destruction of the (2e)-supercurrent when the
transverse magnetic flux through one rhombus is exactly half a
superconducting flux quantum. This destructive interference is
reminiscent of the localization effect predicted for non
interacting charges in \cite{Vidal_98} and considered
experimentally in both superconducting networks \cite{Abilio_99}
and quantum wires \cite{Naud_01}. Interestingly, a finite
supercurrent carried by correlated pairs of Cooper pairs (carrying
a charge of 4e) was predicted to subsist in the case of Josephson
junctions with small capacitances\cite{Doucot_02}.

In experimentally relevant situations, the junctions' capacitances are larger than the ground capacitances of the islands between the junctions. The supercurrent flowing in a linear chain was predicted to be dramatically suppressed, even in chains of rather strong
Josephson junctions\cite{Matveev_02}, because of the large
probability of quantum phase slip events along the chain. As a
result, it is expected that the supercurrent through a phase-biased rhombi chain should be exponentially small.

I. Protopopov and M. Feigelman\cite{Protopopov_04,Protopopov_06}
have studied the equilibrium supercurrent in frustrated rhombi
chains. They have made quantitative predictions for the magnitude
of both, the 2e and the 4e supercurrents, as a function of the
relevant practical parameters : magnetic flux, ratio of Josephson
to Coulomb energy, chain length and quenched disorder. Recently S.
Gladchenko \emph{et al.} reported on the first observation of the
coherent transport of pairs of Cooper pairs in a small size rhombi
array in the quantum regime\cite{Gershenson}.

Whether the chain is in the classical or in the quantum regime is
set by the ratio between the Josephson energy
$E_{J}=i_{c}\frac{\hbar}{2e}$ and the charging energy
$E_{C}=\frac{e^2}{2C}$ of the junctions. In this paper we present
measurements of the current phase relation for long Josephson
junctions rhombi chains ($N=8$ rhombi), carried out either in the
classical phase regime with the Josephson energy much larger than
the charge energy, $E_{J}\gg E_{C}$, or in the quantum phase
regime where $E_{J}/E_{C}\approx 2$. In order to measure the
current phase relation, we shunted the rhombi chain with a high
critical current Josephson junction and measured its switching
current as a function of the magnetic flux for different rhombus
frustrations.

In chapter II we present the theory describing the states and the energy bands for a rhombi chain in the classical limit. This theory is used later in chapter V in order to understand the measurements of the current-phase relation in classical chains. 
In chapter III we begin by a general overview of the phenomena occuring in the presence of charging effects. Secondly we present a theoretical description of quantum fluctuations based on a tight binding hamiltonian for the non frustrated regime. Chapter IV presents the sample
fabrication and characterization. Chapter V is devoted to the
current phase relation measurements in the classical regime where
$E_{J}/E_{C}\approx 20$. These results can be understood from the
shape of the lowest energy band, whose periodicity changes, as
expected, from $h/2e$ at small frustration to $h/4e$ near full
frustration. The corresponding measurements in the quantum limit
for $E_{J}/E_{C}\approx 2$ as well as a detailed quantitative
comparison to the theory are presented in chapter VI. Finally, in the
appendix, we analyze the current voltage characteristics of open
chains where the total phase is not constrained.

\section{Classical Energy states of rhombi chains}

\begin{figure}[htbp]
\includegraphics[width=18cm]{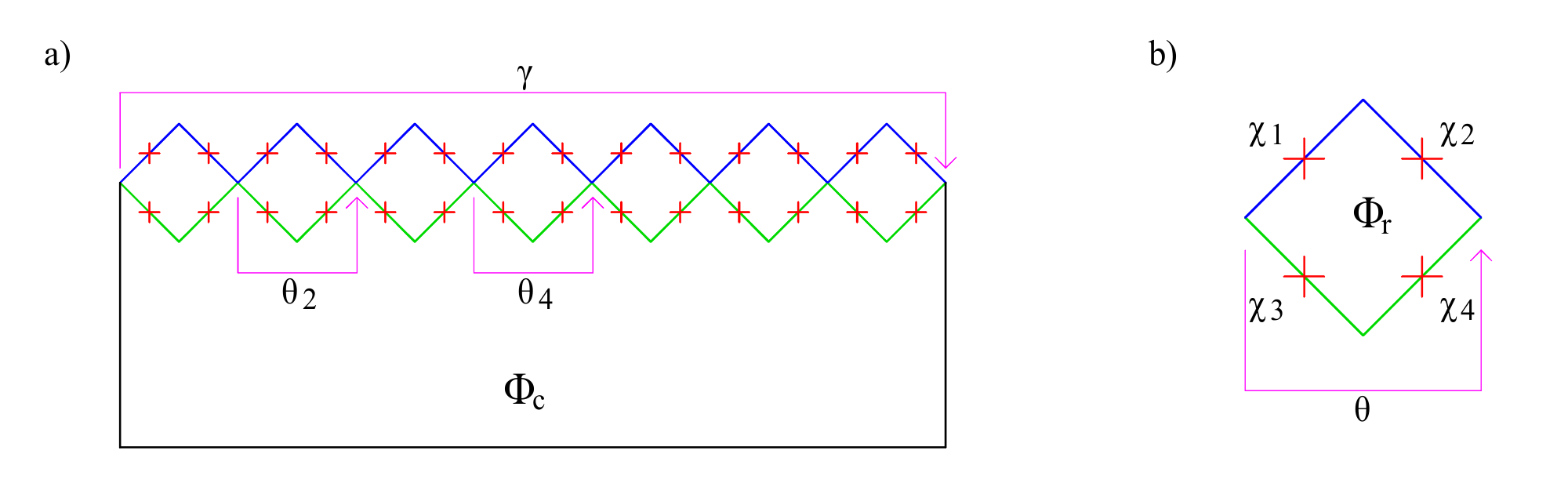}
\caption{(Color online) The different notations used in the text
for the rhombi chain (a) and for one rhombus (b). Each red cross
represents a Josephson junction. The lines represent
superconducting wires and the arrows represent gage invariant
phase differences. The magnetic flux $\Phi_{c}$ inside the
superconducting ring containing the rhombi chain fixes the
phase difference across the chain to $\gamma=2\pi\Phi_{c}/\Phi_{0}$. 
The phase differences over each of the four junctions in one rhombus are denoted by $\chi_{n}$ where $n=1,2,3,4$. The gage-invariant phase $\theta$ will be referred as
the diagonal phase difference. $\Phi_{r}$ represents the magnetic
flux inside one rhombus and the frustration parameter of the
rhombus is given by $f=\Phi_{r}/\Phi_{0}$.} \label{Notations}
\end{figure}

We are interested in the current phase relation $I_{S}(\gamma)$ of a
rhombi chain for different rhombus frustrations $f=\Phi_{r}/\Phi_{0}$. $\Phi_{r}$ represents the magnetic flux inside one rhombus and $\Phi_{0}=\frac{h}{2e}$ is the superconducting flux quantum. The phase difference $\gamma $ over the chain is fixed by introducing the rhombi chain into a superconducting loop
threaded by a magnetic flux $\Phi_{c}=\Phi_{0}\gamma/2\pi$. The
Josephson junctions circuit and the notations that we will further
refer to are represented in Fig.\ref{Notations}.

In this chapter we discuss the case where charging effects are
negligible, and therefore the superconducting phase is a classical
variable. The classical states of one rhombus which depend on the
diagonal phase difference $\theta$ and on the frustration $f$ are
introduced in section A. In section B we extend the classical
description of the energy states to a chain containing N rhombi.
In this case the energy band depends again on the frustration $f$
and the phase difference $\gamma$ over the whole chain.

\subsection{Single Rhombus}

We consider a single rhombus made of 4 identical Josephson
junctions (Fig.\ref{Notations}b) with Josephson energy $E_{J}$ and
critical current $i_{c}=\frac{2e}{\hbar}E_{J}$.
\begin{figure}[htbp]
\begin{center}
\begin{minipage}[c]{8.5cm}
a) \\
\includegraphics*[width=7.5cm]{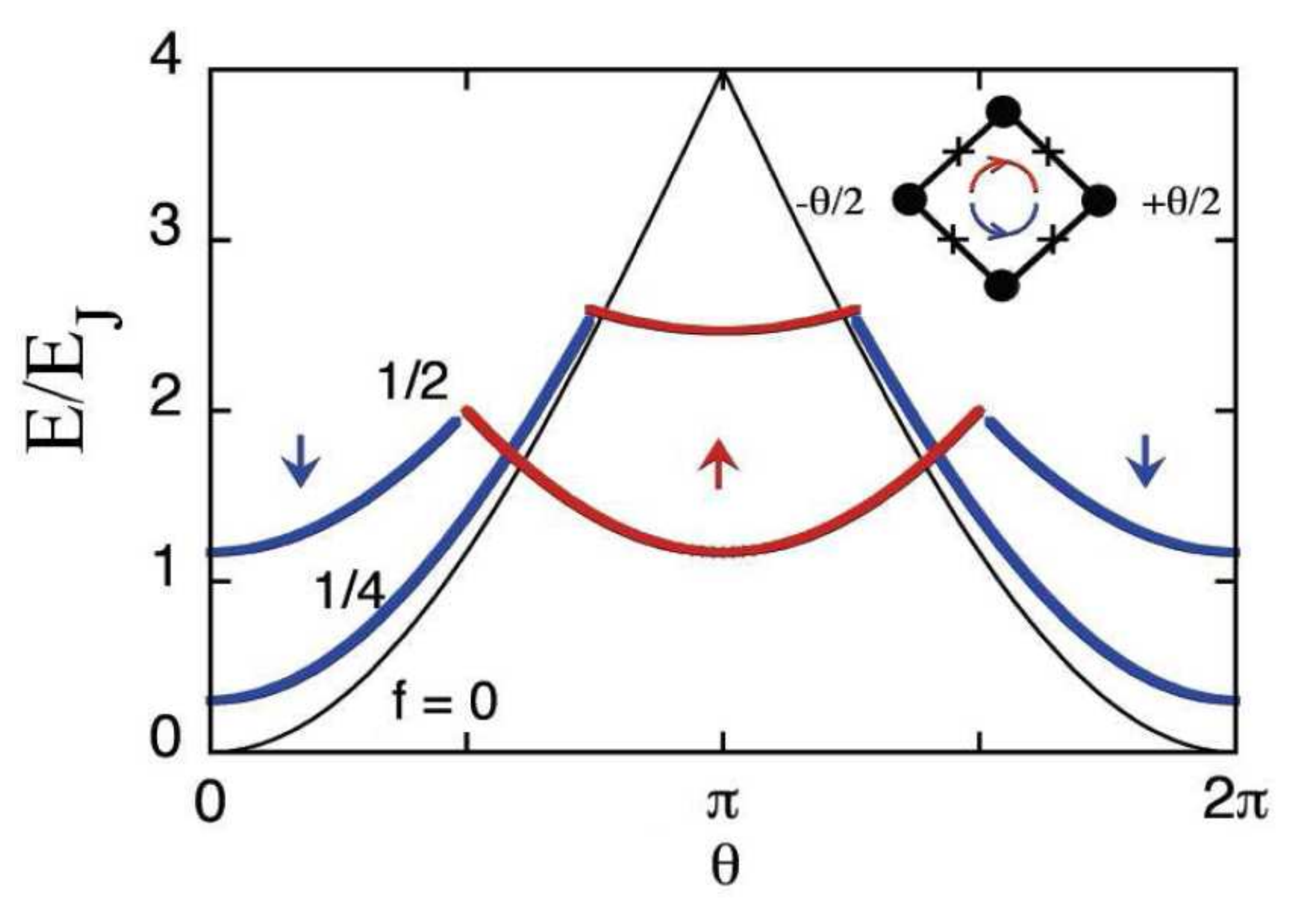}
\end{minipage}
\hfill
\begin{minipage}[c]{8.5cm}
b) \\
\includegraphics*[width=8cm]{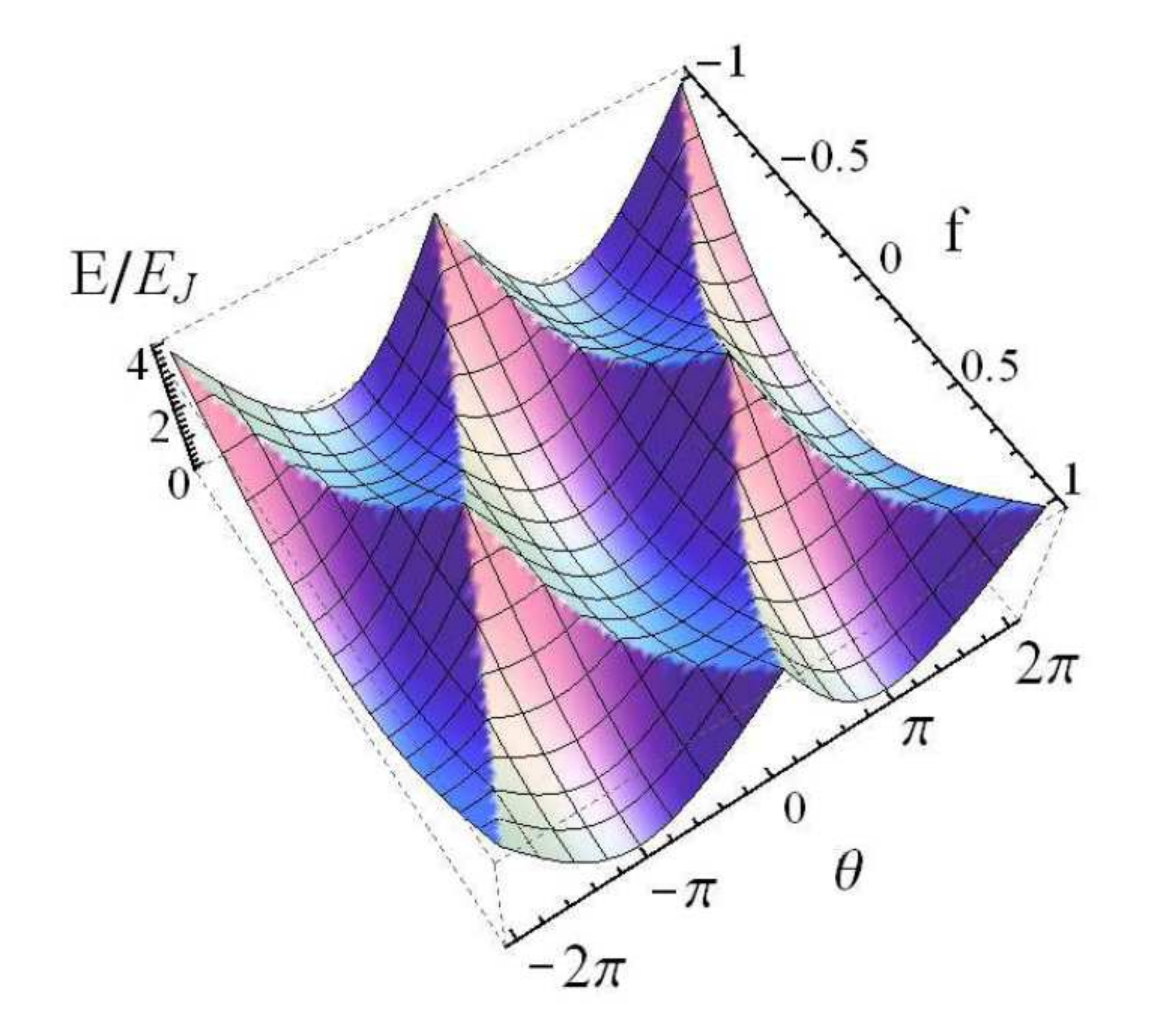}
\end{minipage}

\vspace{0.5cm}
\begin{minipage}[c]{8.5cm}
c) \\
\includegraphics*[width=7cm]{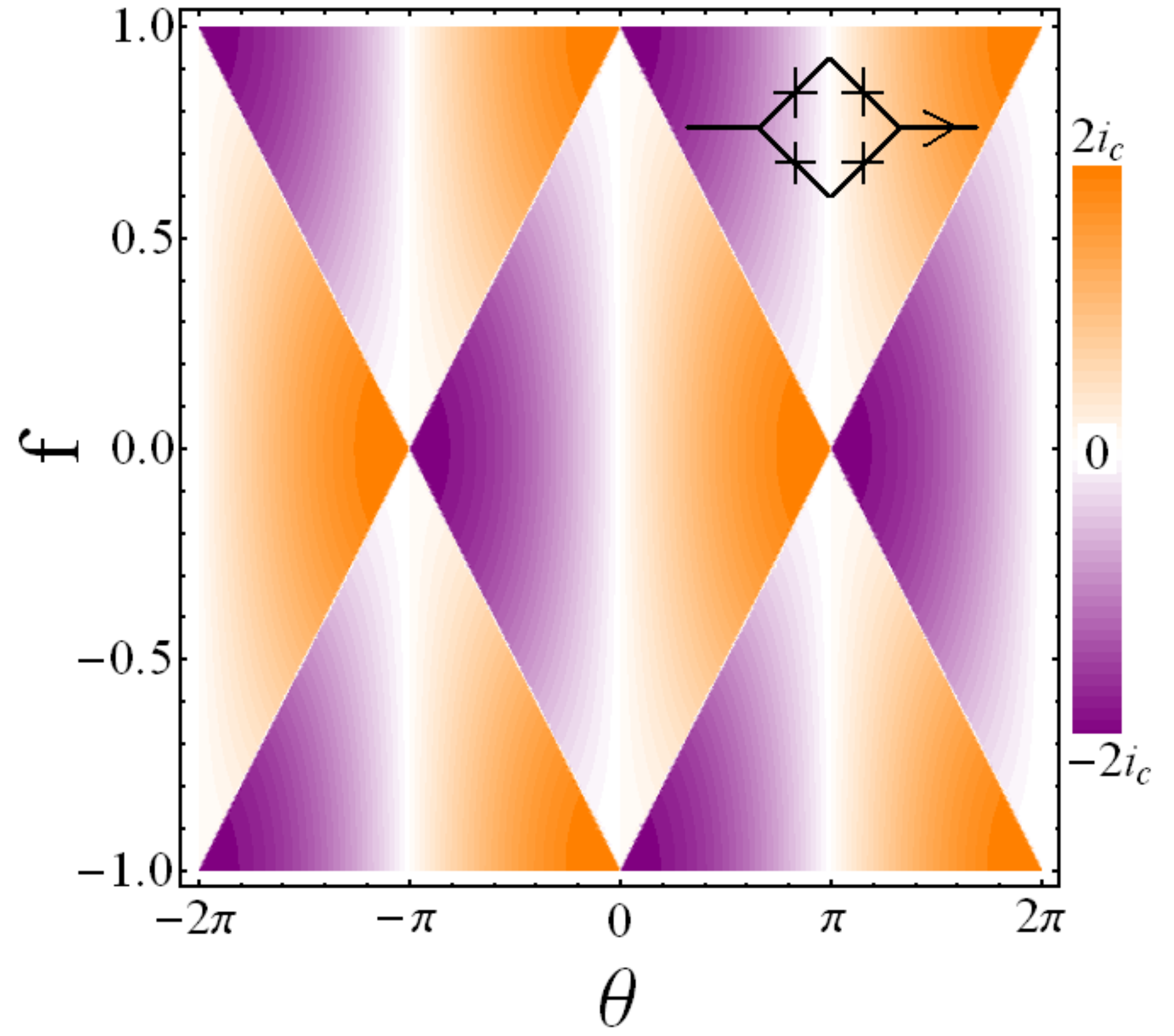}
\end{minipage}
\hfill
\begin{minipage}[c]{8.5cm}
d) \\
\includegraphics*[width=7cm]{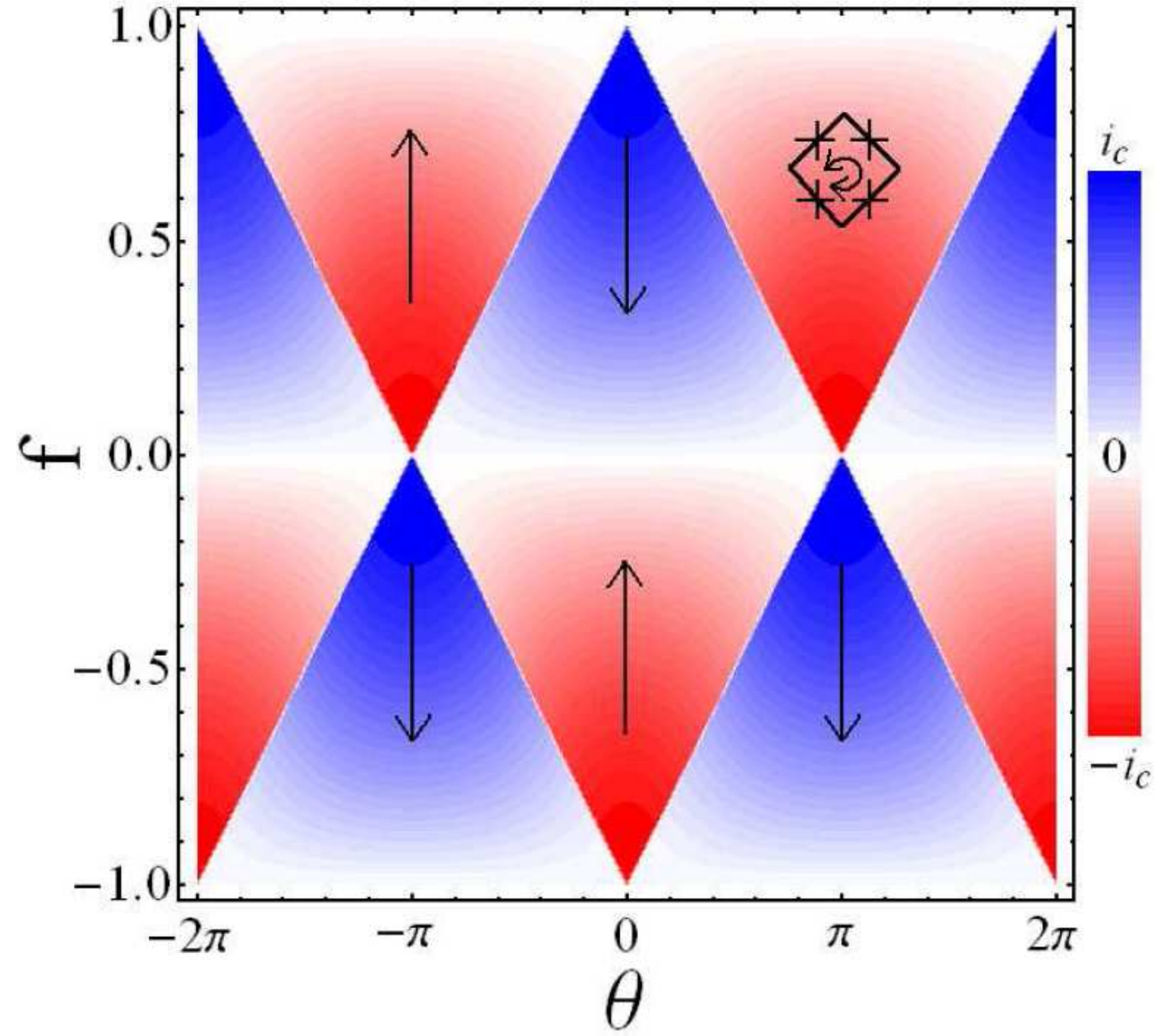}
\end{minipage}
\end{center}

\caption{(Color online) Classical states of a single rhombus. a) The ground state
energy as a function of the diagonal phase difference $\theta$
over the rhombus for three different frustrations $f=0$, $1/4$ and $1/2$. The inset
shows the two possible persistent current states: a clockwise
flowing supercurrent (blue lines) and a
counterclockwise supercurrent (red line). b) 3D plot
showing the lowest energy band as a function of $\theta$ and
$f$. c) Two dimensional plot for the supercurrent
across one rhombus. The amplitude and sign of the supercurrent is
illustrated by the background color: orange(positive values) for currents flowing
from left to right and violet(negative values) for currents from right to left. d)
Two dimensional plot for the amplitude and the direction of the
persistent current around the ring. The clockwise current states,
denoted $\left|\downarrow\right\rangle$, are represented in blue, the
counterclockwise current states, denoted $\left|\uparrow\right\rangle$, in
red. At full frustration ($f=0.5$) the ground state is
degenerate for $\theta=\pm \pi/2$, and the two eigenstates differ by
the sign of the persistent current.} \label{rhombus}

\end{figure}

Neglecting additional terms due to inductances, the potential
energy of one rhombus containing four identical junctions, is
simply given by the sum of the Josephson energies of the four
junctions:
\begin{equation}
\label{ERhombiClassique}
E(\chi_{1},\chi_{2},\chi_{3},\chi_{4})=\sum_{n=1}^{4} E_{J}(1-
\cos\chi_{n})
\end{equation}
The sum of the phases $\chi_{n}$ is fixed by the flux inside the
rhombus:  
\begin{equation}
\label{PhaseCondition}
\sum \chi_{n} = 2\pi f
\end{equation}
Using the notations defined earlier, the ground state energy of one rhombus, in the classical
regime ($E_{J}\gg E_{C}$), is found by minimizing the energy
(\ref{ERhombiClassique}) and depends on the parameters $\theta$
and $f$:
\begin{equation}
\label{ERhombiClassique2} E(\theta,f)/E_{J}=4-2
(\left|\cos(\theta/2+\pi f/2)\right|+\left|\cos(\theta/2-\pi
f/2\right|)
\end{equation}

A complete description of the phase diagram for one rhombus is
given in Fig.\ref{rhombus}. The circular current in the
superconducting ring is $i_{p}(\theta,f)=\frac{2e}{\hbar}
\frac{\partial E(\theta,f)}{\partial f}$ and it is $2\pi$-periodic
in $\theta$ and f (Fig.\ref{rhombus}d). The supercurrent
through one rhombus is given by $i_{s}(\theta,f)=\frac{2e}{\hbar}
\frac{\partial E(\theta,f)}{\partial \theta}$ and it is shown in
Fig.\ref{rhombus}c.

The interesting feature about this system is the change from
$2\pi$ to $\pi$ periodicity as a function of the bias phase
$\theta$ over the rhombus when the frustration f changes from $0$
to $1/2$. This property does not exist in the case of a dc SQUID, as there is no modulation of the energy as a function of $\theta$ at full frustration. At $f=1/2$ the rhombus has two classical ground states, $\theta=0,\pi\pmod {2\pi}$, denoted in analogy to the
z-projection of the spin $\frac{1}{2}$ by $\left|\downarrow\right\rangle$ and
$\left|\uparrow\right\rangle$ respectively. These two states have the same
energy $E(\theta=0,f=0.5)=E(\theta=\pi,f=0.5)=2(2-\sqrt{2})E_{J}$
but opposite persistent currents (see Fig.\ref{rhombus}d). In the case of
a current biased rhombus, the phase $\theta$ is controlled via the
current phase relation of a single rhombus
$i_{s}(\theta,f)=\frac{2e}{\hbar} \frac{\partial
E(\theta,f)}{\partial \theta}$ by the external current. The
critical current of a single rhombus is given by the maximum
supercurrent through the rhombus for a given frustration $f$:
$I_{c}=\max(i_{s}(\theta))_{f=const}=\max(\frac{2e}{\hbar} \frac{\partial
E(\theta,f)}{\partial \theta})_{f=const}$. It is periodic in $f$ and
varies from a maximum of $2i_{c}$ down to $i_{c}$. For $-1/2\leq f
\leq 1/2$ it reads :
\begin{equation}
\label{icrhombus}
I_{c}=2  i_{c} \cos^{2}\frac{ \pi f}{2}
\end{equation}
\subsection{Rhombi chain}

In order to understand the classical states of the chain we can start our
analysis with the case where each rhombus has a well defined diagonal phase difference
across it. For a closed chain of $N$ identical rhombi, the sum of
all the diagonal phase differences $\theta_{n}$ is fixed by the
magnetic flux $\Phi_{c}$ to a total phase difference $\gamma$ over
the chain (see Fig.\ref{Notations}a).
\begin{equation}
\label{phasechain}
\sum_{n=1}^{N}\theta_{n}=\gamma
\end{equation}

In the region where the frustration, $0\le f \ll1$, is small we
obtain by minimizing the total energy that the diagonal phase
differences over each rhombus are identical up to a constant
multiple of $2\pi$. The phase difference across the diagonal of
the $n$-th rhombus in the state $|m\rangle$ is given by:

\begin{equation}
  \theta_n=\frac{\gamma-2\pi m}{N}+2\pi m_n\ \; , \;   m=\sum_{n}m_n
  \label{theta}
\end{equation}
where $m$ is the number of vortices inside the superconducting
loop that contains the rhombi chain, and $m_n$ is an integer
corresponding to the number of vortices that crossed the $n$-th
rhombus. Therefore the ground energy of the chain is $N$ times the
energy of a single rhombus:
\begin{equation}
\label{EChain} E(\gamma,f)/E_{J}=N(4-2 (\left|\cos((\gamma-2\pi
m)/2N+\pi f/2)\right|+\left|\cos((\gamma-2\pi m)/2N-\pi
f/2\right|)
\end{equation}

At $f=0$ and in the limit $N>>1$ the expression above can be
developed around zero. The energies of the low-lying states are
given by:
\begin{equation}
  E_{m}(\gamma)= \frac{E_J}{2N}\left(\gamma-2\pi m\right)^2
\label{expansion}
\end{equation}

\begin{figure}[htbp]
\begin{center}

\includegraphics[width=8.5cm]{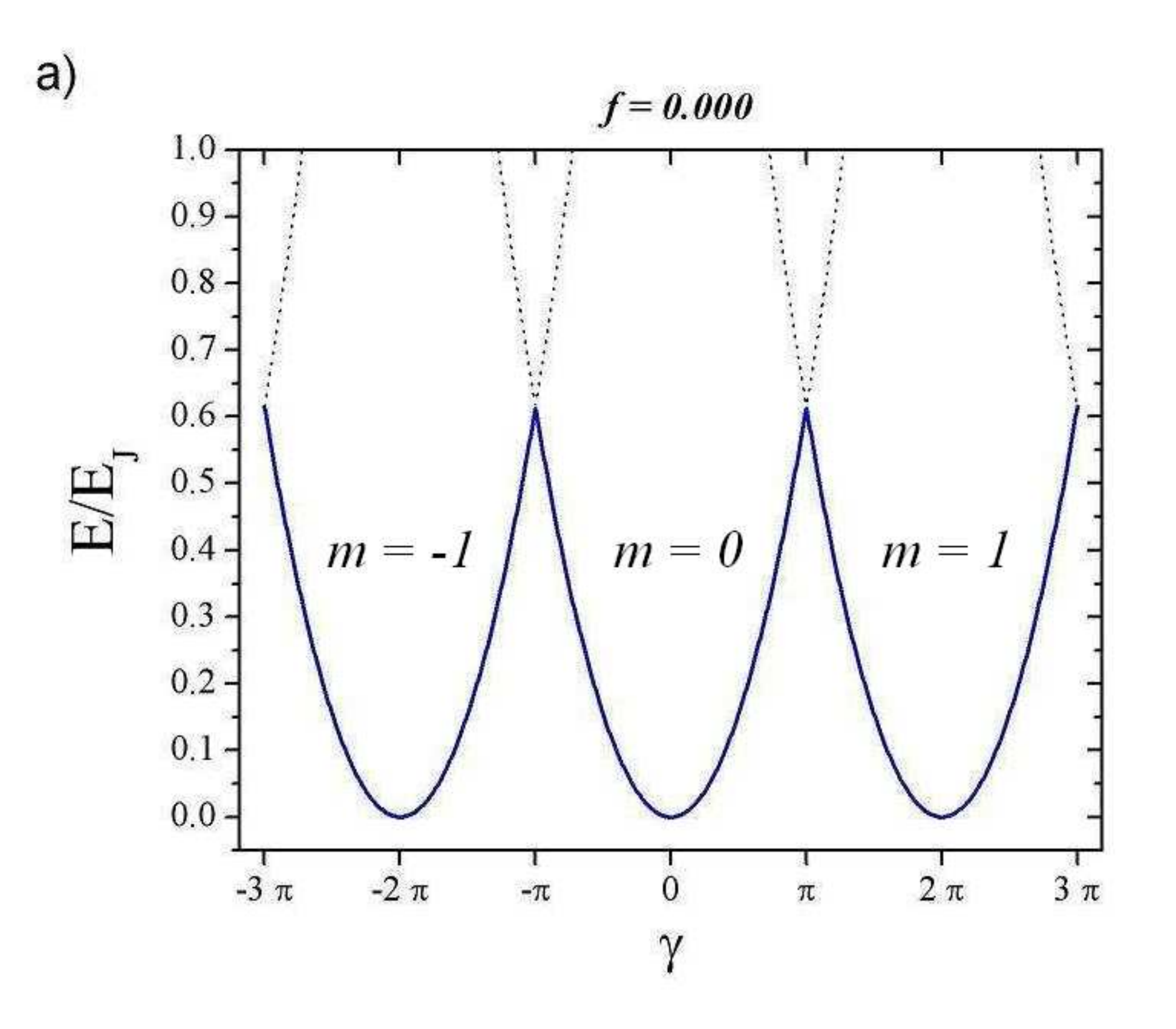}
\includegraphics[width=8.5cm]{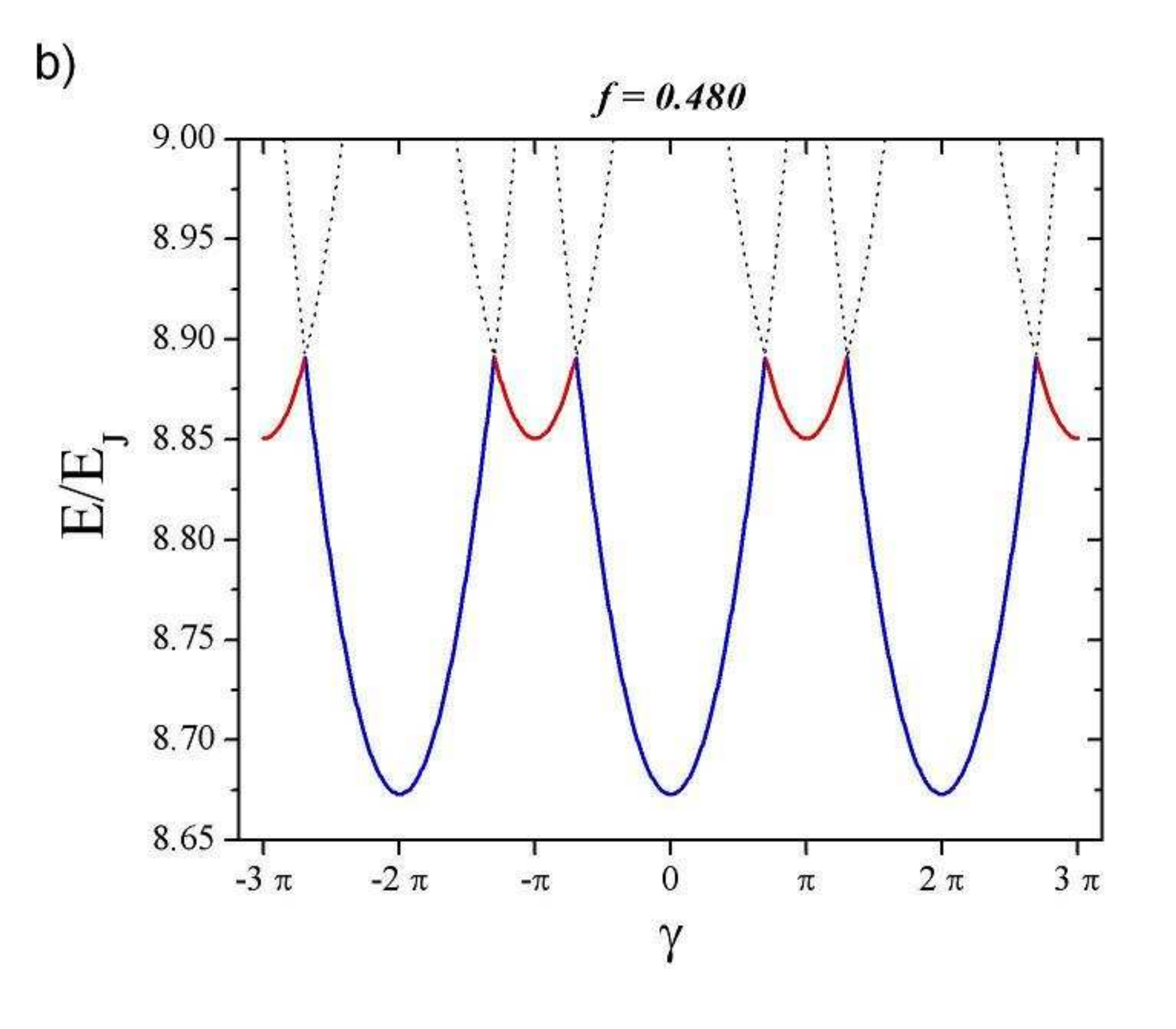}
\includegraphics[width=8.5cm]{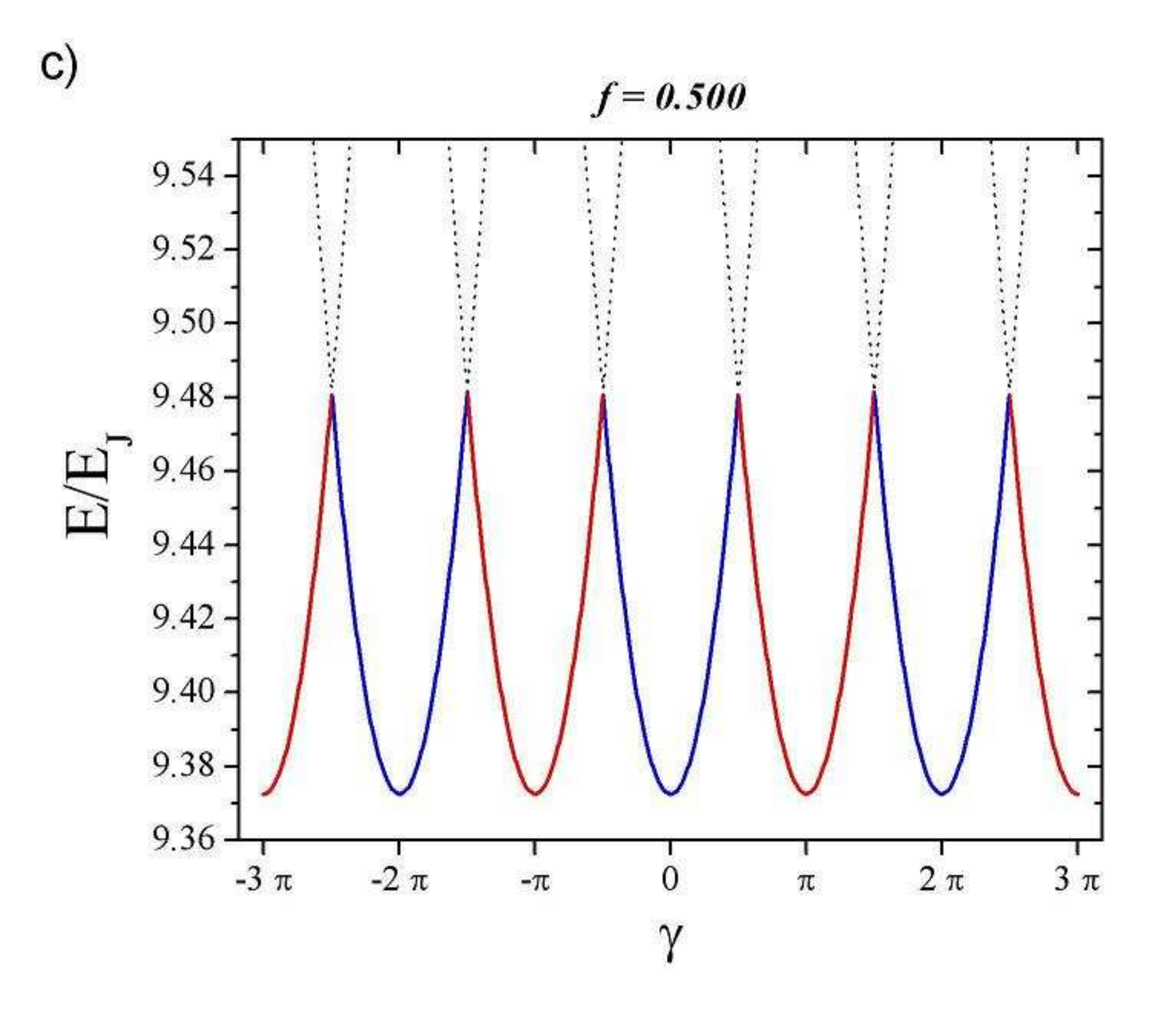}
\includegraphics[width=8.5cm]{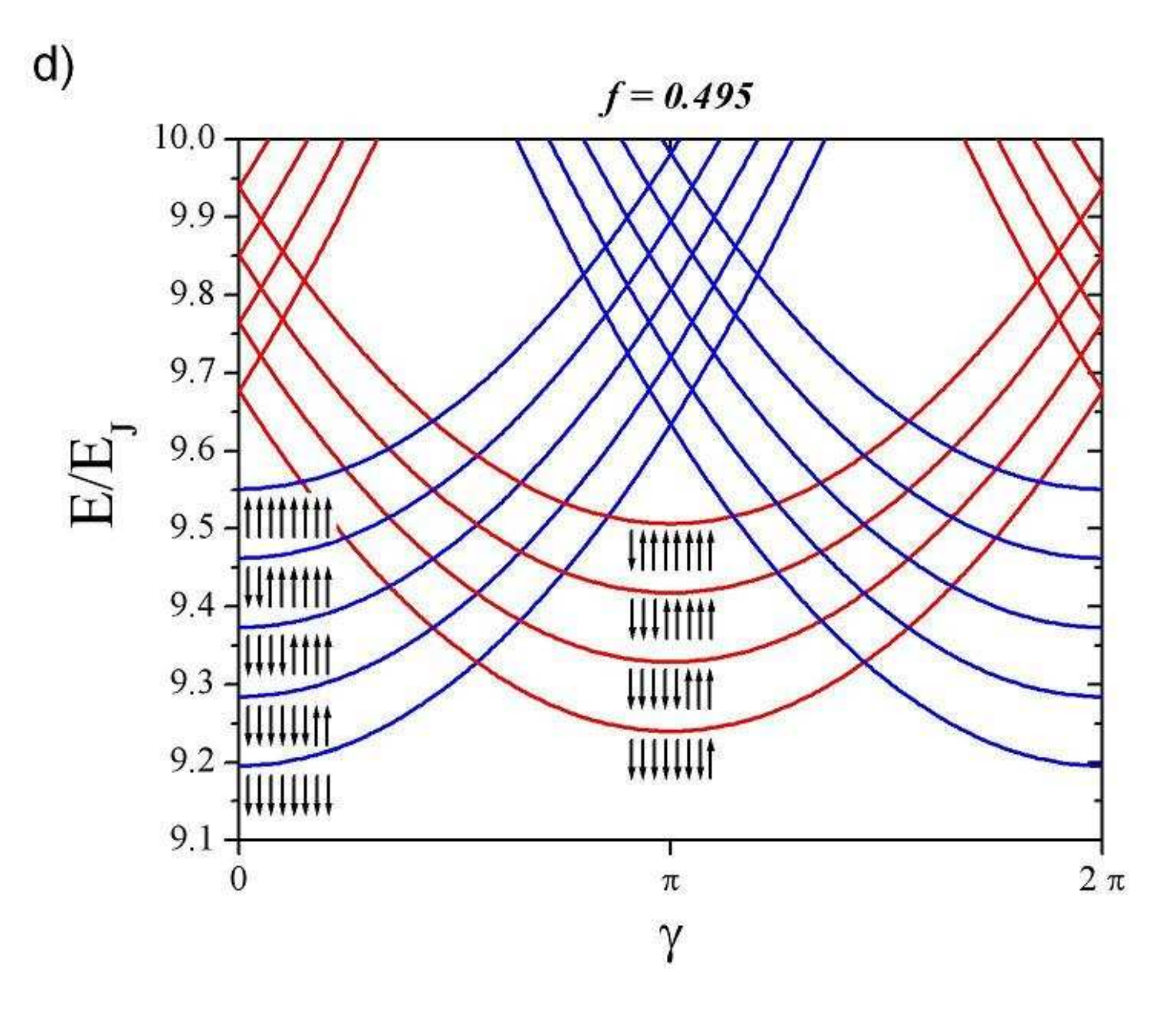}

\end{center}

\caption{(Color online) a) Ground state energy of a classical 8
rhombi chain as a function of the phase $\gamma$, for zero (a),
intermediate (b) and full frustration (c). The plot colors
correspond to the $\left|\downarrow\right\rangle$ state (blue) and
$\left|\uparrow\right\rangle$ state (red). The supercurrent
flowing through the chain is given by the derivative of the energy
as a function of $\gamma$ and consists of a series of unequal
sawtooth in the vicinity of $f=1/2$.  d) Identification of the
lowest energy states of the chain near full frustration
($f=0.495$). The up and down arrows indicate the spin states of
the $8$ rhombi. Note the change in the parity for the number of
switched rhombi between successive minima.}

\label{chainstates}

\end{figure}

The ground state energy consists of a series of shifted arcs, with
period $2\pi$ as shown in Fig.\ref{chainstates}a.
In analogy to a single rhombus, at small frustration all rhombi of
the chain are in the $\left|\downarrow\right\rangle$ state (see
Fig.\ref{rhombus}). The supercurrent through the chain is given
by the derivative of the ground state energy with respect to
$\gamma$, $I_S(\gamma,f)= \frac{2e}{\hbar}\frac{\partial E(\gamma,f)}{\partial
\gamma}$. Therefore the current-phase relation of an unfrustrated chain, in the classical
regime, is a $2\pi$-periodic sawtooth function as for a single
rhombus. But in contrast to a single rhombus the critical current
of a chain with large $N$ is approximatively $N$ times smaller than
the critical current of a single junction. The value for the
critical current of the chain $i_{c}\frac{\pi}{N}$ can be easily calculated
from the energy expansion (\ref{expansion}).

As $f$ approaches $1/2$, the total energy can be reduced by
flipping the spin state of one rhombus. The chain with $N-1$ rhombi in the
$\left|\downarrow\right\rangle$ state and one rhombus in the
$\left|\uparrow\right\rangle$ state becomes energetically more
favorable near $\gamma=\pi\pmod{2\pi}$ as shown in Fig.\ref{chainstates}a. 
Thus the energy diagram consists of an
alternate sequence of arcs, centered respectively at even and odd
multiples of $\pi$. At full frustration $f=1/2$, the period as a
function of $\gamma$ turns to $\pi$(Fig.\ref{chainstates}a upper
trace). Here, the energy modulation
$\frac{\pi^{2}}{8N\sqrt{2}}E_{J}$ and the maximum supercurrent
$I_{s}=i_{c}\frac{\pi}{2\sqrt{2}N}$ are significantly weaker than
at zero frustration. The crossoverpoint between these two regimes
is defined as the minimum frustration that induces at $\gamma=\pi$
a flip from the $\left|\downarrow\right\rangle$ state to the
$\left|\uparrow\right\rangle$ state for one single rhombus in the
chain. In Fig.\ref{chainstates}b we represented the state of
the system for $f$ slightly larger than the crossover frustration.
For large $N$ the width of the frustration window scales with
$1/N$ and can be approximated by the condition :
\begin{equation}
\label{window} 1 -\tan \frac{\pi f}{2} < \frac{\pi^{2}}{8N}
\end{equation}

Within this window, the supercurrent is expected to show a complex
sawtooth variation as a function of the phase $\gamma$ with unequal
current steps. It is interesting to discuss in more details the
structure of the chain states in the vicinity of the full
frustration region. In \cite{Protopopov_04} it has been shown that
near $f=0.5$ the energy of the different possible chain states can
be approximated by the formula:
\begin{equation}
\label{ec_chainstates} E_{m,S^z}(\gamma)\approx
\frac{E_J\sqrt{2}}{4N}(\gamma+N\pi/2+\pi S^z -2\pi
m)^2-\sqrt{2} \delta S^z E_J +const
\end{equation}
where $\delta=2\pi f-\pi$. Here $S^z=-\frac{1}{2}\sum$sign$
  (\sin(\theta_n))$ corresponds to the z-projection of the total spin
S describing the whole rhombi chain. Figure \ref{chainstates}d
shows the energy diagram for the lowest energy chain states with
N=8 in order to highlight the topological distinctions between branches with minimas at
even and odd values of $\gamma/\pi$. Near $\gamma=0$ the ground
state is obtained when all the rhombi are in the
$\left|\downarrow\right\rangle$ state. Near the next minimum, one rhombus has
flipped into the $\left|\uparrow\right\rangle$ state. For the higher energy
levels one can conclude in general that at even values of
$\gamma/\pi$, chain states containing an even number of rhombi in
the $\left|\uparrow\right\rangle$ state (so called even states) show a
minimum. At odd values of $\gamma/\pi$ chain states with an odd
number of rhombi in the $\left|\uparrow\right\rangle$ state (so called odd
states) show a minimum. At full frustration $f=1/2$ all chain
states with an even and odd number of flipped rhombi become
respectively degenerate. Complete degeneracy is achieved at full
frustration at $\gamma=\pi/2$ : even and odd states have the same energy.\\

In conclusion, the current phase relation of the rhombi chain in
the classical regime should follow a sawtooth like function with a
slowly varying amplitude as a function of the frustration except
for a small region around $f=0.5$. Inside this so called
frustration window the periodicity of the sawtooth should double
and its amplitude should drop by a factor of $2\sqrt{2}$. In
chapter V we present measurements that precisely confirm these
predictions.
\\

\section{Quantum Energy states of rhombi chains}

In this chapter we will discuss the influence of charging effects
on the current phase relation of the rhombi chain. Section A
offers a qualitative overview of the expected phenomena when
quantum fluctuations are not negligible. Section B is devoted to a
quantitative analysis in the region $f=0$. We develop a tight
binding model proposed initially by Matveev et al\cite{Matveev_02}
for a Josephson junctions chain. This theoretical model will
successfully fit our measurements presented in section VI.

\subsection{Quantum phase slips}

\begin{figure}
\includegraphics[width=15cm]{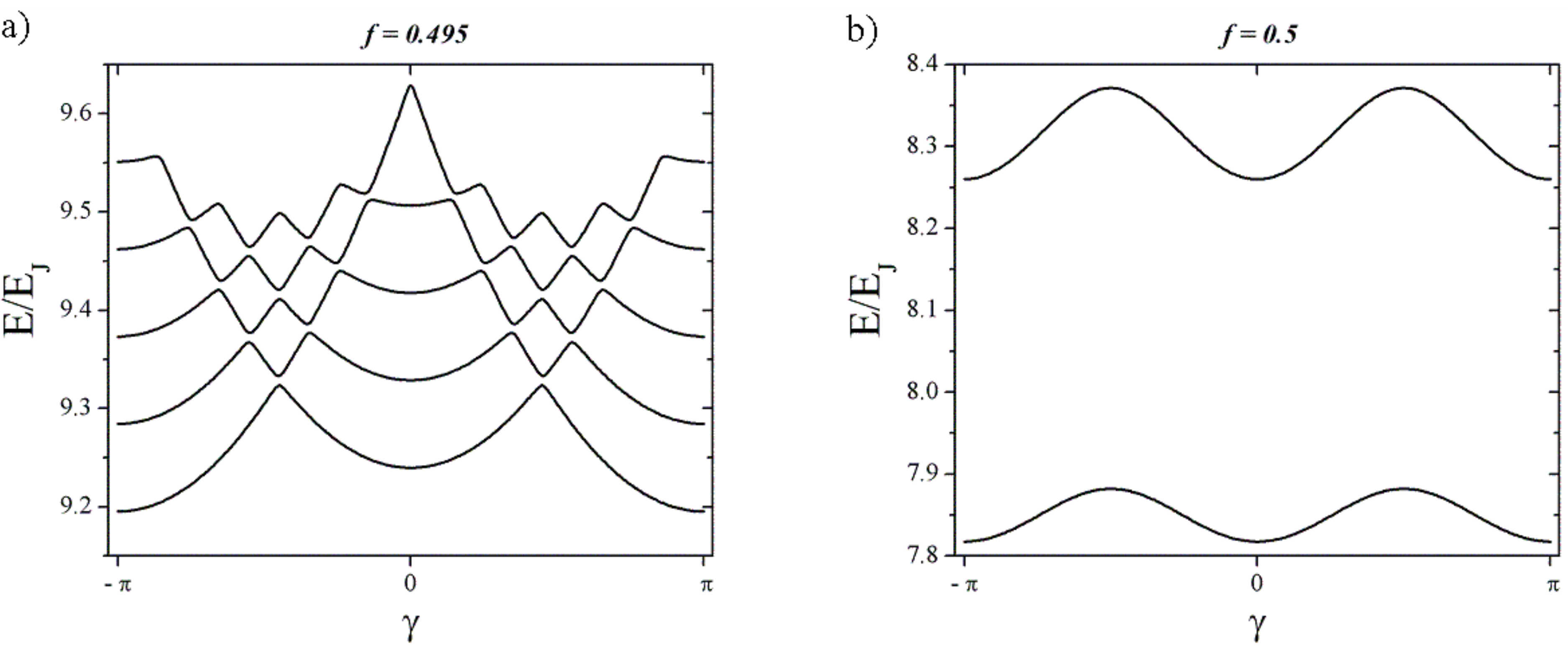}
\caption{Energy bands of the rhombi chain in the presence of quantum fluctuations. a) Energy bands of an almost classical chain near full frustration ($f=0.495$). In this regime quantum phase slips mix the degenerate classical levels and lead to the appearance of avoided level crossings. b) First two energy bands of the quantum chain at full frustration ($f=0.5$). Deep in the quantum regime fluctuations mix a lot of different classical levels and the spectrum is no longer piecewise parabolic. The precise form of the spectrum can be calculated on the basis of the Eq. (16) of the reference \cite{Protopopov_04}. Figure b) was generated by taking $E_J/E_C=6$. } 
\label{QuantumEnergy}
\end{figure}

Quantum fluctuations start to play a role when the charging energy
$E_{C}$ cannot be neglected anymore in comparison to the
Josephson energy $E_{J}$. Quantum fluctuations induce quantum
phase slips. For quantum junctions at very low temperature the
role of quantum phase slips is twofold. First, phase slip events,
even rare, allow the system to tunnel through the energy barriers
which separate the local minimums and to reach the ground states
discussed above. On the other hand phase slips induce quantum
coupling between different states and lead to the formation of
macroscopic quantum states extended over the whole
chain\cite{Doucot_02,Matveev_02,Protopopov_04}. This superposition
of states lifts the high degeneracy of the classical states. In
the case of important quantum fluctuations, the crossing points
between different states shown in Fig.\ref{chainstates} become
anticrossing points, strongly modifying the physical properties of
the system (see Fig.\ref{QuantumEnergy}). The rate of phase slips depends on the height and
shape of the energy barrier which is set by the ratio $E_{J}/E_{C}$. We choose to focus
our attention on two extreme cases: $E_{J}\gg E_{C}$ (classical
regime) where there are practically no phase fluctuations and
$E_{J}/E_{C}\approx  2$ (quantum regime) where the quantum
fluctuations open a significant gap between the classical states
at the crossing points.
\\

The frustrated and non frustrated regime involve different kinds
of tunnelling process:

At \emph{$f=0$} (see Fig. \ref{chainstates}a) or  when
$f$ is outside the window defined by equation \ref{window}, the
energy states cross each other at $\gamma=\pi$ (modulo $2\pi$).
The necessary $2\pi$ jump can be achieved by simultaneous phase
slips events in two junctions of one rhombus (one junction in each
branch). At $f=0$, the simplest path corresponds to a sinusoidal
energy barrier of $4E_{J}$ as shown in Fig.\ref{2PiFlip}. The
rhombi chain can be treated like a Josephson junctions chain
considered by Matveev et al. \cite{Matveev_02}, except that, here,
the tunnel amplitude for quantum phase slips ($v$) involves the
simultaneous phase slip on two junctions. We have calculated this
tunnel amplitude in the case of a rhombus and the next section
presents a detailed description of the tight binding model that we
used to fit our experimental results in the quantum regime at
$f=0$.

Qualitatively, when quantum fluctuations are large enough, one
expects a rounding of the sawtooth-like $2e$ supercurrent
turning eventually to a sinusoidal current of exponentially small
amplitude, as predicted in \cite{Matveev_02}. At finite
frustration, the tunnel path is flux dependent and involves more
complex trajectories in the multidimensional energy landscape. The
tunnel amplitude will presumably be increased.
Near \emph{$f=1/2$,} the successive energy minima as a function
of $\gamma$ have periodicity $\pi$. The corresponding chain states
differ by the sign of the persistent current in one rhombus. Here the transition
requires a phase jump of $\pi$ across one rhombus. The energy
barrier for this process can be approximated by considering a path
in the parameter space where a single junction switches by $2\pi$.
In this case the energy barrier is close to a sinusoidal barrier
with height $2(\sqrt2-1)E_{J}$, $i.e.$ $0.414$ times the
energy barrier for a single junction.\\

\subsection{ Quantum fluctuations of the rhombi chain at zero
magnetic field}

In the region where the frustration is small, $0\le f \ll1$, the
theory we develop here is just a slight modification of the
analysis  carried out in  \cite{Matveev_02} for a chain of
single Josephson junctions. The reason for this similarity is that around
zero frustration the energy of a single rhombus as a function of
the phase difference across it  has only {\it one} minimum (see
Fig.\ref{rhombus}a). This implies  the coincidence of the
classification of the classical states for our system and for the single Josephson junctions chain. In this section we present the theory of quantum
fluctuations in a non-frustrated rhombi chain which we used to fit
the experimental data. In our analysis we assume that the
Josephson energy of the junctions is much larger than the charging
energy and quantum fluctuations in {\it individual} Josephson
junctions are small. However, as we will see below, the
fluctuations in the whole chain can be strong.

\begin{figure}
\includegraphics[width=300pt]{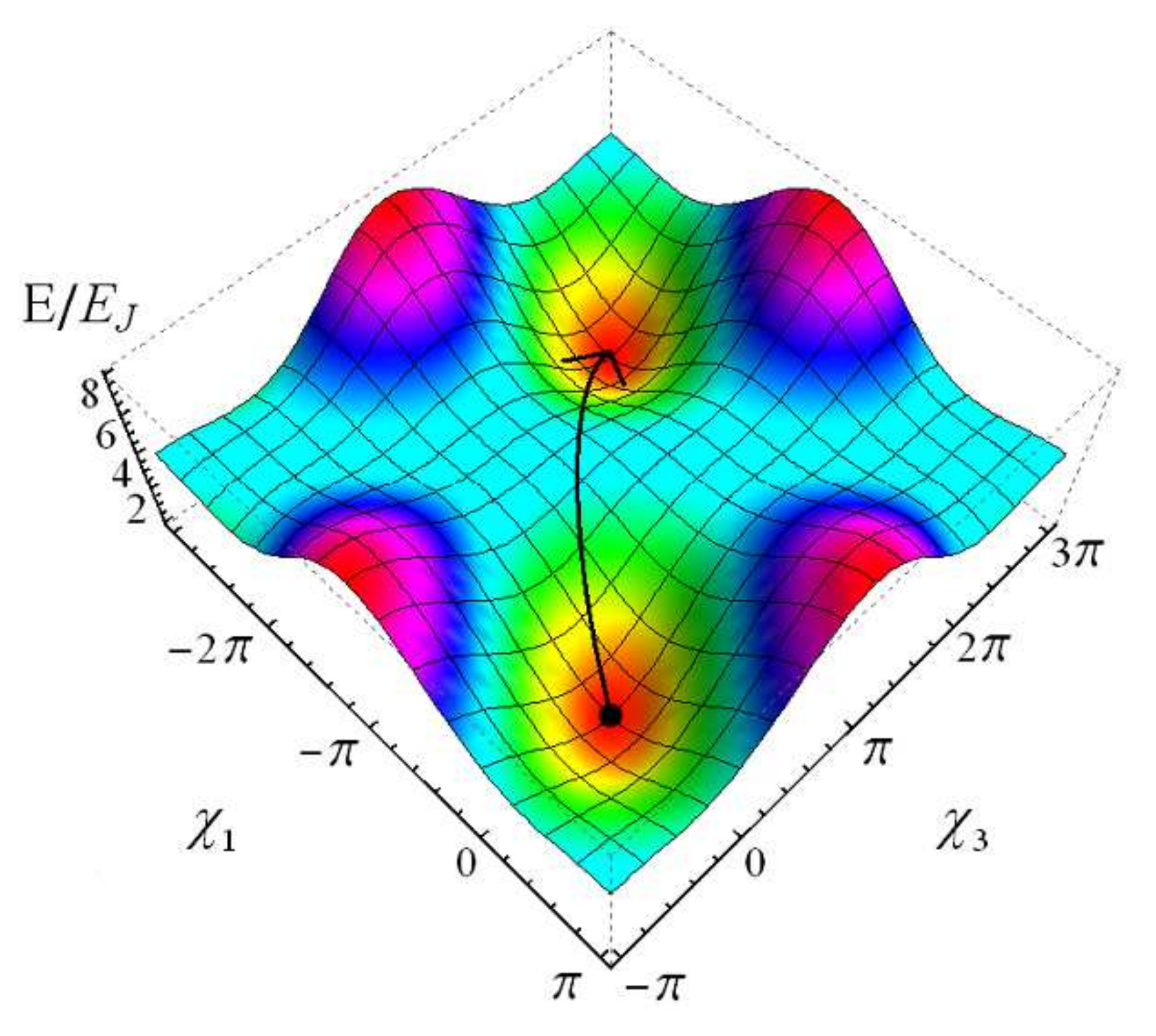}
\caption{(Color online) The energy landscape for one rhombus at
zero frustration ($f=0$) as a function of the phase differences
$\chi_{1}$ and $\chi_{3}$ of Josephson junctions in opposite
branches of the rhombus. The transition of the chain from state
$\left|m\right\rangle$ to state $\left|m+1\right\rangle$
corresponds to a $2\pi$ jump of one of the diagonal rhombus phases
$\theta_n$. This transition implies a simultaneous flip of the two
phases $\chi_{1}$ and $\chi_{3}$ respectively by $-2\pi$ and
$2\pi$. The arrow in the figure represents the corresponding
classical trajectory. The phenomenon can also be seen as the
process of one vortex crossing the rhombus ring.} \label{2PiFlip}
\end{figure}

Quantum fluctuations (more precisely quantum phase slips) lead to the mixing of
classical states described above. At large $E_J/E_C$ this effect
can be described within the tight-binding approximation (cf.
\cite{Matveev_02}). Classical states
lie far from each other in the configuration space and are
separated by barriers of the order $E_J$ (see Fig.\ref{2PiFlip}). At large $E_J/E_C$
the amplitude of quantum tunnelling from state
$\left|m\right\rangle$ to $\left|m'\right\rangle$ is exponentially
small and decreases fast with the increase of the distance between
$\left|m\right\rangle$ and $\left|m'\right\rangle$. For a given
state $\left|m\right\rangle$ the closest states in the
configuration space are $\left|m\pm1\right\rangle$. To achieve the
state $\left|m+1\right\rangle$ one needs to change the phase
difference across the diagonal of one rhombus by $2\pi$ (at large
$N$, cf. eq. (\ref{theta})). Since we need to maintain the sum of
the phase differences around the rhombus (fixed by the zero flux
inside it, see eq. (\ref{PhaseCondition})) we need to change by $\pm2\pi$ the phase differences
over {\it two} junctions in different branches of the rhombus (see Fig.\ref{2PiFlip}). Let us denote the amplitude of such a process by
$\upsilon$. In a semiclassical approximation this amplitude is
determined by the vicinity of the classical trajectory connecting
states $\left|m\right\rangle$ and $\left|m+1\right\rangle$ in
imaginary time
\begin{equation}
  \upsilon=A \exp\left(-S_0\right)
  \label{upsilon}
\end{equation}
Here $S_0$ is the imaginary-time action on the classical
trajectory (instanton). As it is easy to see from the preceding
discussion, $S_0$ is just twice the action describing a phase slip
in a single junction. We thus have (cf. equation 7 of the
reference \cite{Matveev_02}, note the difference in the
definitions of $E_C$ in this paper and in \cite{Matveev_02})
\begin{equation}
  S_0=2\sqrt{\frac{8E_J}{E_C}}
  \label{S_0}
\end{equation}
The coefficient $A$ in (\ref{upsilon}) accounts for the
contribution of the trajectories close to the classical one.
Standard calculation gives
\begin{equation}
  A\approx 4.50 (E_J^3 E_C)^{1/4}
  \label{A}
\end{equation}

We can now construct the tight-binding Hamiltonian describing the
effect of the phase slips on the properties of the chain
\begin{equation}
  H|m\rangle=E_m|m\rangle +4N\upsilon|m+1\rangle+4N\upsilon|m-
  1\rangle
  \label{Ham}
\end{equation}
The coefficient $4$ in the total tunneling matrix element  is due
to the number of possible tunneling paths within one rhombus while
$N$ appears here because of the fact that a phase slip in any
rhombus brings
 the system to the same state.

Following now the procedure described in \cite{Matveev_02} we can
reduce the problem of finding the eigenvalues of the Hamiltonian
(\ref{Ham}) to the solution of the Mathieu equation
\begin{equation}
  \psi''(x)+(a-2q\cos2x)\psi(x)
  =0\,, \qquad   \psi(x+\pi)=e^{i\gamma}\psi(x)
  \label{Mat}
\end{equation}
The parameters of the Mathieu equation are defined by
\begin{equation}
  a=\frac{2NE}{\pi^2 E_J}\,, \qquad q=\frac{8 N^2 \upsilon}{\pi^2 E_J}
\end{equation}
Here $E$ is the energy of the rhombi chain.

The equation (\ref{Mat}) can be solved analytically in different
limiting cases (see ref. \cite{Matveev_02} for details). By solving
it numerically and using the general relation
$I_{S}=\frac{2e}{\hbar}dE/d\gamma$ one can find the current-phase
relation for the rhombi chain at arbitrary fluctuations' strength. This is the exact procedure that we have used in chapter VI in order to fit the measured current-phase relation for quantum chains. We found a very good agreement between the theoretical predictions and the measured data.

\section{Sample fabrication and characterization}
The samples were made by standard e-beam lithography and shadow
evaporation technique using a Raith Elphy Plus e-beam
system\cite{Raith} and an ultra high vacuum evaporation chamber.
They consist of small arrays of $Al/AlO_{x}/Al$ tunnel junctions
deposited on oxidized silicon substrates.
The respective thicknesses of the Al layers were 20 and 30 nm.The tunnel barrier
oxidation was achieved in pure oxygen at pressures around $10^{-3}
mbar$ during 3 to 5 minutes depending on the sample. The samples
were mounted in a portable closed copper block which was thermally
anchored to the cold plate of either a $He^{3}$ insert or a
dilution fridge. All lines were heavily filtered by thermocoaxial
lines and $\pi-$filters integrated in the low temperature copper
block. Additional low frequency noise filters were placed at the
top of the cryostat.

\begin{figure}[htbp]
\centering
\includegraphics[width=12cm]{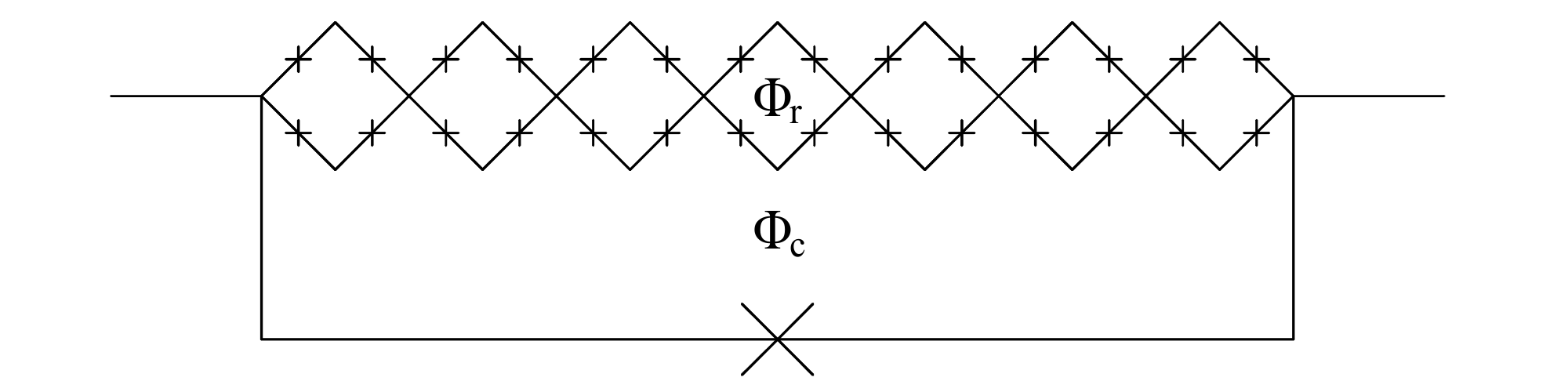}
\caption{Schematic of the circuit designed to measure the current phase relation
in a rhombi chain. The chain is closed by a superconducting line interrupted by
an additional Josephson junction referred to as the "shunt junction". An external
magnetic field $B$ allows the control of both the rhombi frustration $f$
and the total phase of the closed chain $\gamma$.}
\label{schematic}
\end{figure}
In order to measure the current-phase relation, we introduced the
rhombi chain in a closed superconducting loop which contains an
additional shunt Josephson junction as shown in
Fig.\ref{schematic}. We have measured the switching current of this circuit. The switching current was obtained from the switching histogram. We fixed the threshold voltage at about one third of the shunt junction gap voltage. The histograms were accumulated at a rate $10kHz$ using a fast trigger circuit\cite{trigger}. The bias current was automatically reset to zero immediately after each switching event. The switching current $I_{SW}$ corresponds, in our definition, to an escape probability of $50\%$.

As the critical current of the shunt junction is much larger than the critical current of the chain, near the switching event the phase difference over it is close to $\pi/2$. Therefore the flux $\Phi_c$ changes only the phase difference $\gamma$ over the chain. The switching current through the parallel circuit represented in Fig.\ref{schematic} can be written as the sum of the partial supercurrents in the two branches. 
\begin{equation}
\label{currentsum} I_{SW}=I_{S}(\gamma-\frac{\pi}{2}) + I_{c}\sin(\frac{\pi}{2})
\end{equation}
Here, $I_{S}(\gamma)$ is the supercurrent in the rhombi chain and
$I_{c}$ is the shunt junction critical current. Therefore the
$\gamma$ dependence of the switching current of the shunted
rhombi chain directly reflects the current phase relation of the
rhombi chain.

The frustration inside the rhombi chain was
controlled by a constant external perpendicular magnetic field.
The flux inside the closed chain could be either applied
simultaneously or swept independently using control lines. In the
first case the two parameters $\gamma$ and $f$ are linked by
the area ratio between the rhombus and the ring (see Table \ref{sampleswitch}). Since the rhombus area is much smaller than the ring area, using small variations of the magnetic field $B$ we can vary the phase $\gamma$ for an approximately constant value of $f$.

To be able to achieve reversible fine tuning of the phases, we found it crucial to avoid any flux trapping in the vicinity of the superconducting circuit. For this purpose the
superconducting leads were patterned with linear open voids which
ensure free motion of vortices. Different sample designs were
investigated including open and closed chains. The typical
elementary junction area ranged from $0.15 \times 0.15 \mu m^{2}$
to $0.3 \times 0.6 \mu m^{2}$. The Josephson energy was inferred
from the experimental tunnel resistance of  individual junctions
and the nominal Coulomb energy was estimated from the junction
area using the standard capacitance value of $50 fF/\mu m^{2}$ for
aluminum junctions. In general, the measured area of
the junctions was slightly smaller than expected. The actual
Coulomb energy is therefore larger (by about $20\%$) than its
nominal value.

\begin{figure}[htbp]
\centering
\includegraphics[width=9cm]{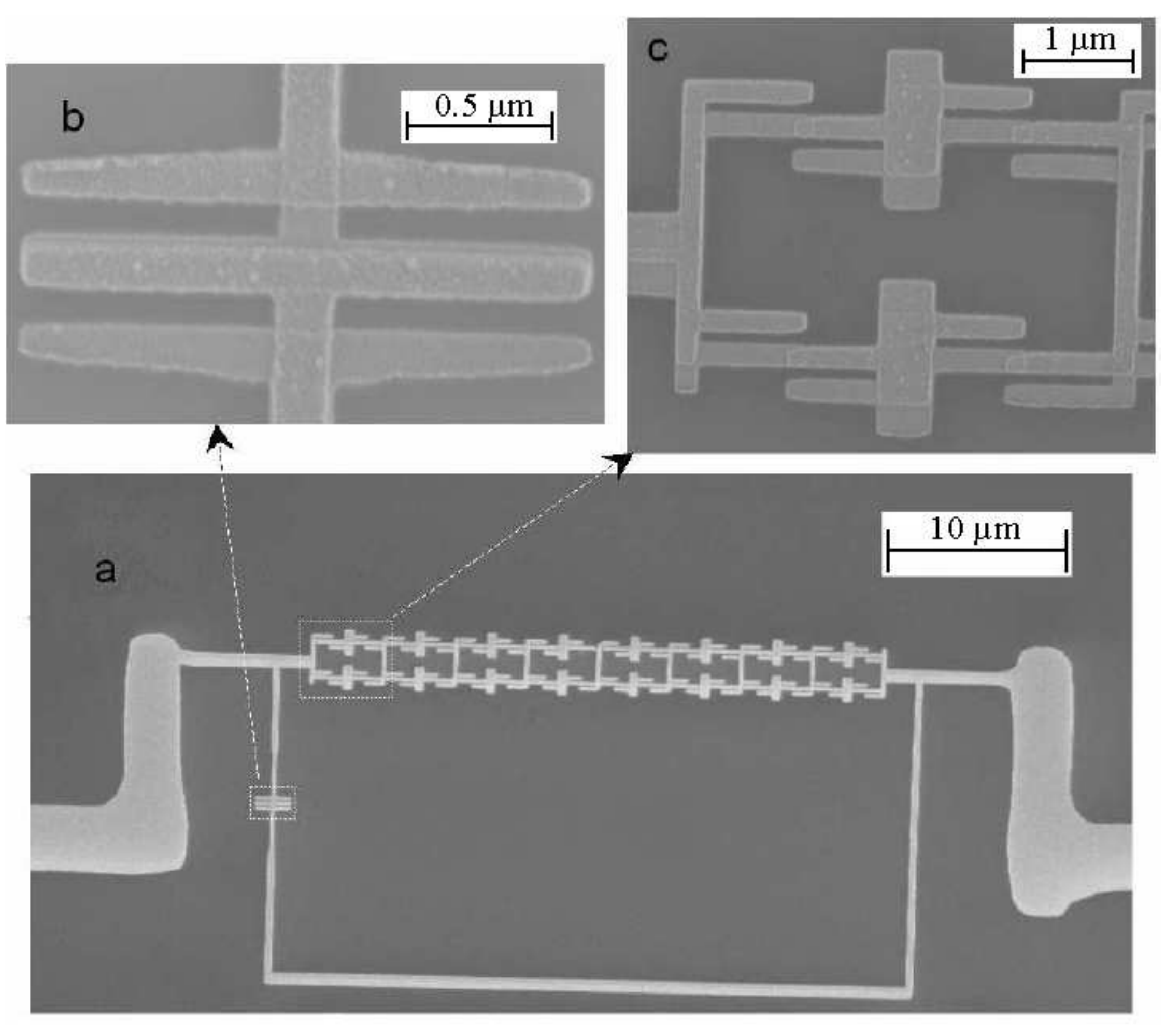}
\caption{a) SEM image of a rhombi chain ($N=8$, sample C) in the closed superconducting
circuit. The shunt junction is visible on the left vertical line. b) An enlarged
image of one rhombi is also presented. c) For small magnetic field variations, the flux
inside the rhombi practically remains constant, while the total phase on the array varies.}
\label{sample}
\end{figure}

 We designed, for this experiment, a series of samples as shown in Fig.\ref{sample}.
$E_{J}$ and $E_{c}$ as well as the number of rhombi were chosen near the range of the
optimum parameters prescribed in Ref \cite{Protopopov_04}.  The shunt junction has a critical current about $10$ times larger than the switching current of the chain.
Fig.\ref{sample}c shows a SEM image of one rhombus. The actual design of the resist
mask was optimized to insure the best homogeneity of junction critical currents\cite{homogeneity}.

We concentrate on results for three particular samples with the following common characteristic
parameters :  number of rhombi $N=8$, rhombus area : $2 \times 4 \mu m^{2}$, shunt junction :
$0.15 \times 2 \mu m^{2}$. Other parameters are listed in Table \ref{sampleswitch}.

\begin{table}[htdp]
\caption{Characteristic parameters of the samples. For sample A, $E_{J}$
was obtained from the tunnel resistance measured on a reference open chain fabricated on the same chip. For sample B, $r_{n}$ (not measured) was estimated to be similar. The charging energy was inferred from the nominal junction area.
$R_{shunt}$ ($\Omega$) is the tunnel resistance of the shunt junction.}
\begin{center}
\begin{tabular}{|c||c|c|c|c|c|c|c|}
\hline
 sample & ring area & rhombi junctions area & $E_{c}(K)$ & $r_{n}(\Omega)$ & $E_{J}(K)$ &  $E_{J}/E_{c}$ & $R_{shunt}$   \\
\hline
A & $8 \times 40 \mu m^{2}$ & $0.15\times0.30\mu m^{2}$ & 0.43 & $850$ & 9.0 & $20$ & $169$ \\
\hline
B  & $18 \times 36 \mu m^{2}$ & $0.15\times 0.30\mu m^{2}$ & 0.43 & $-$ & $-$ & $-$ & $167$  \\
\hline
C  & $18 \times 36 \mu m^{2}$ & $0.15\times 0.15\mu m^{2}$ & 0.8 & $4860$ & 1.6 & $2$ & $627$  \\
\hline
\end{tabular}
\end{center}
\label{sampleswitch}
\end{table}

\section{Classical chains}
\begin{figure}[htbp]
\centering
\includegraphics[width=10cm]{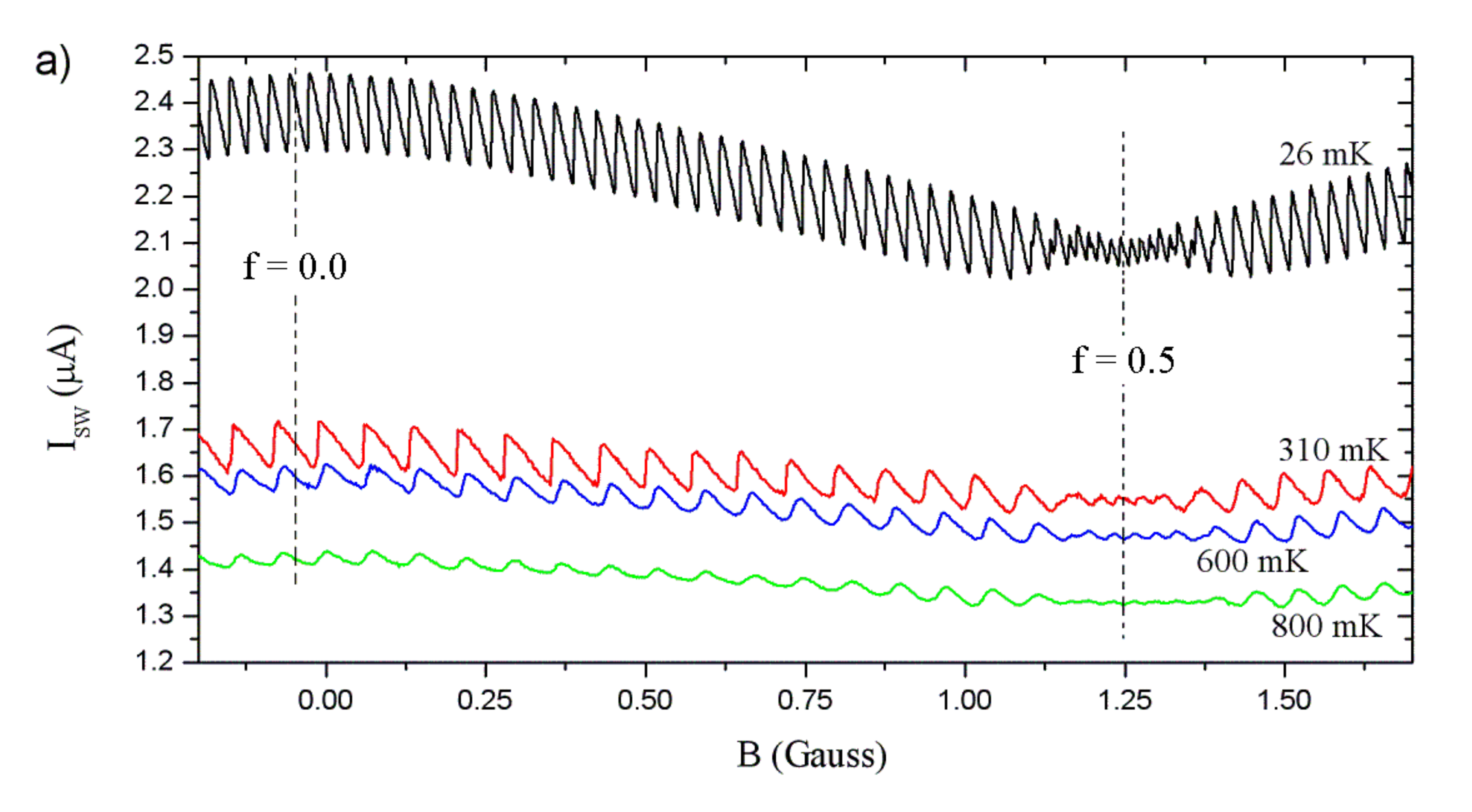}
\includegraphics[width=18cm]{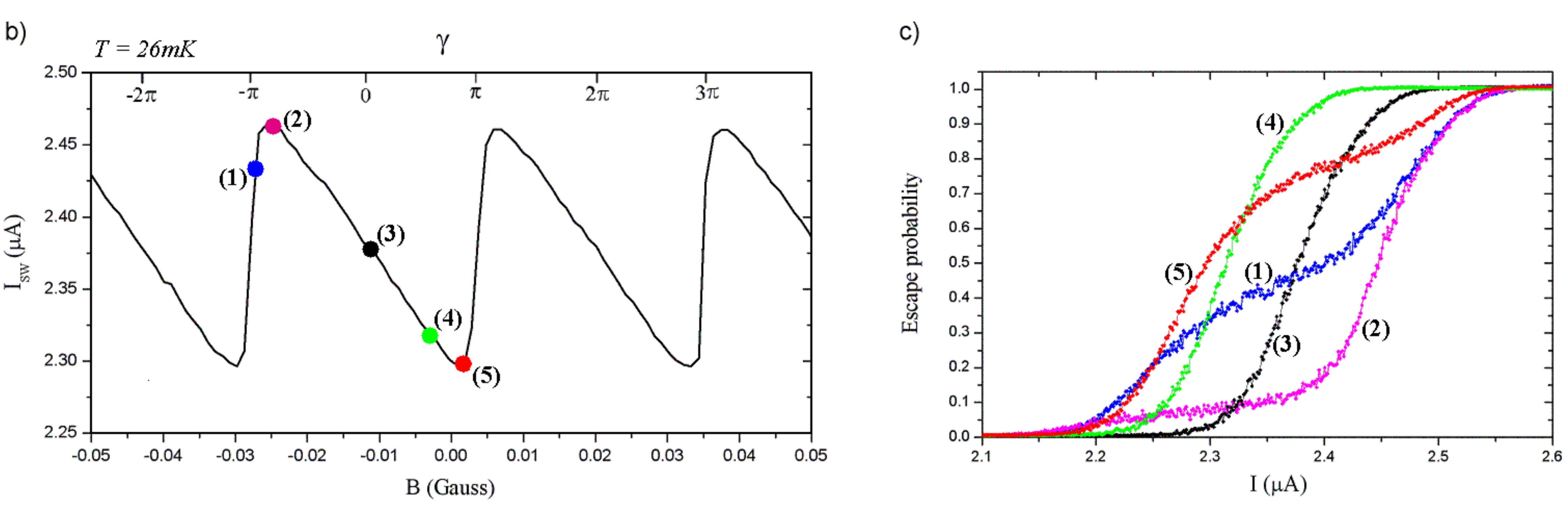}
\caption{(Color online) a) Experimental plot of the switching current $vs$ external magnetic field at different temperatures. The sample B was measured at the lowest temperature $T=26 mK$. The higher temperature measurements correspond to the sample A. b) Magnification of the region near rhombi frustration $f=0$. c) The switching probability of the circuit rhombi chain + shunt junction (sample B) $vs$ current bias at different magnetic fields which correspond to the numbered points indicated in figure \ref{oscillation}b.}
\label{oscillation}
\end{figure}

The observed dependence of the switching current $I_{SW}$ $vs$ the external magnetic flux is shown in Fig.\ref{oscillation}. Both the rhombi frustration $f$ and the phase along the main ring $\gamma$ are controled by the magnetic field. 
We observe a complex dependence of $I_{SW}$ as a function of the magnetic flux with mainly one slow periodic envelop of period $2.57$ Gauss that we attribute to the frustration inside the rhombus and one fast sawtooth oscillation that we understand as the modulation of the supercurrent as a function of the phase $\gamma$. The number of periods differs for the two samples A and B as expected from the difference between the ring areas.

\begin{figure}[htbp]
\centering
\includegraphics[width=9cm]{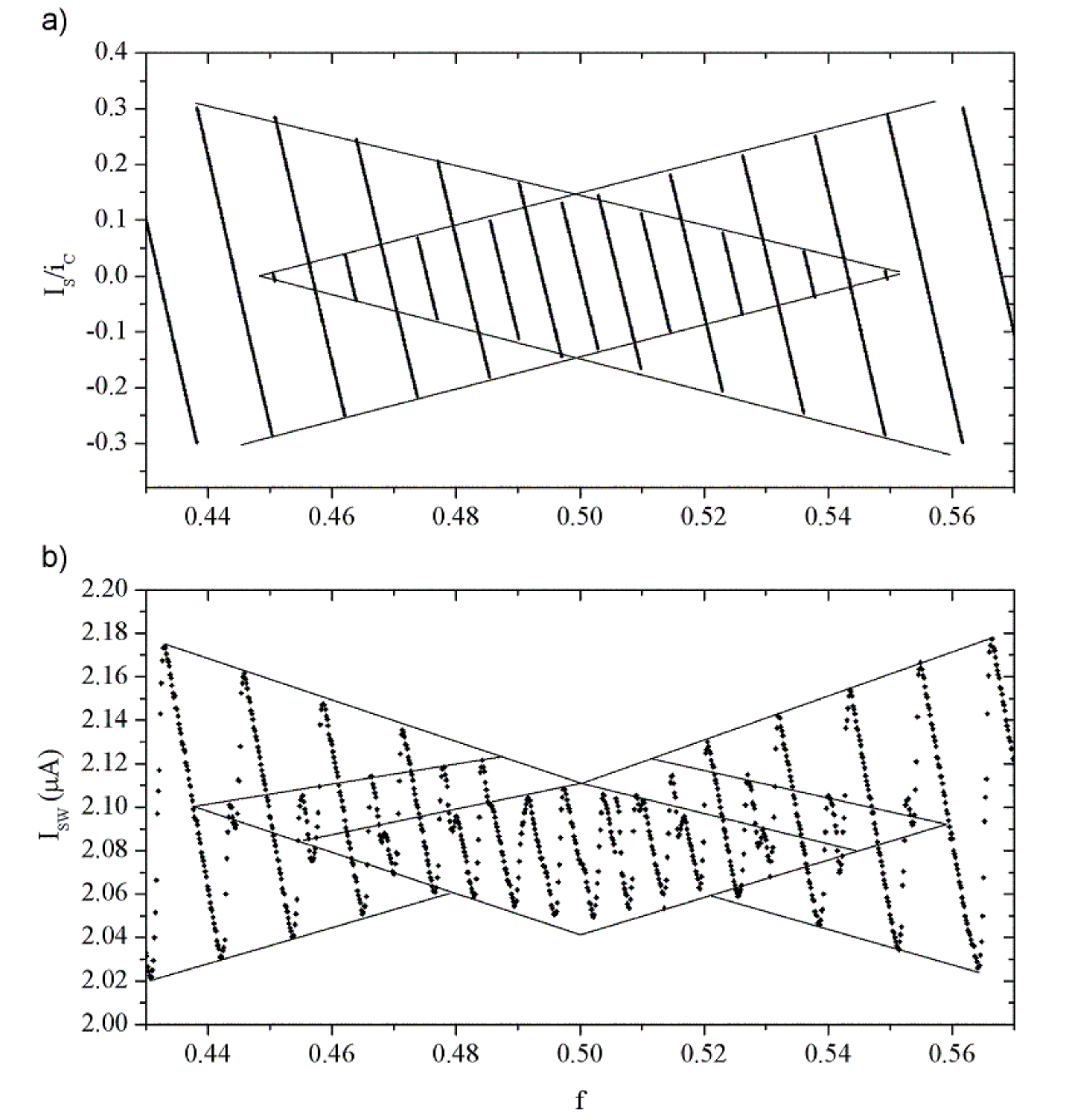}
\caption{Comparison between the  measured switching current
of sample B (b) and the theoretical persistent
current (a) in the classical limit (see section II) near $f=1/2$.
The lines joining the cusps are guides for the eyes. The vertical axis in (a) is in units of the single junction critical current $i_{c}$.}
 \label{sawtooth}
\end{figure}

We have verified that the fast modulation is periodic with period $
h/2e$ except near $f=1/2$ where the period is $h/4e$ (see Fig.\ref{oscillation}a and Fig.\ref{sawtooth}b).
This result confirms precisely what is illustrated in Fig.\ref{chainstates} : the chain states undergo a transition from phase periodicity $2\pi$ to periodicity $\pi$ when the rhombi are fully frustrated. Let us notice here that the half periodicity is not actually visible over many periods since the control magnetic flux changes both the frustration and the phase. Instead, we do observe a sequence of saw teeth with unequal amplitudes which become regular only at exactly $f=1/2$. We have confirmed the period halving in a separate experiment  where we used on-chip superconducting lines to control $f$ and $\gamma$ separately. We could observe up to 12 oscillations (not shown) of the critical current $vs$ $\gamma$ when the rhombi frustration was fixed exactly at $f=1/2$ by a static magnetic field $B=1.29$ gauss.

\begin{figure}[htbp]
\centering
\includegraphics[width=9cm]{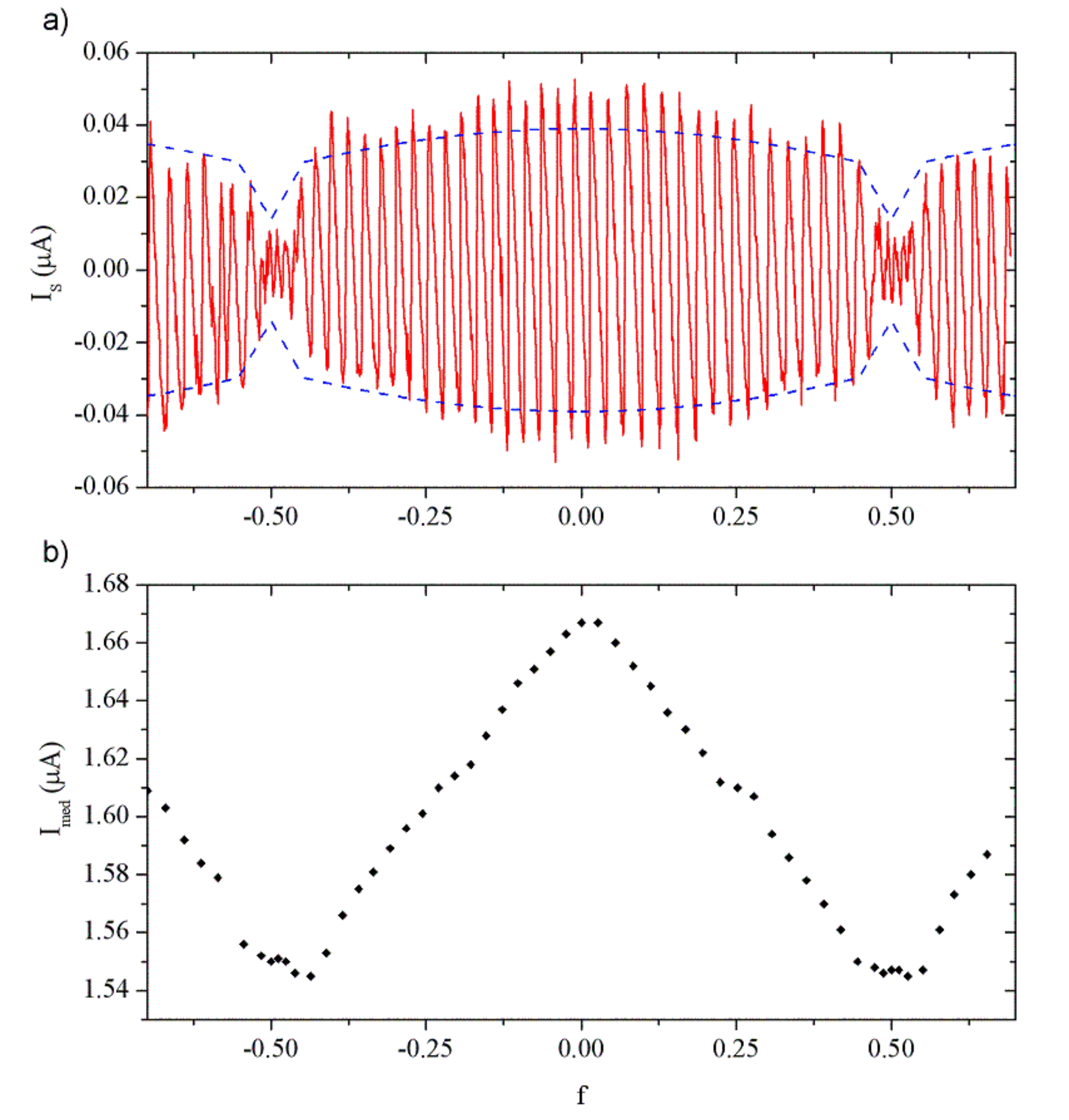}
\caption {(Color online) The fast oscillation component
$I_{s}$ (a) and the median component $I_{med}$ (b) of the switching
current in sample A at $T=310 mK$. The plot (a) represents the
supercurrent through the chain and the plot (b) gives the additional
contributions coming from the shunt junction and the rhombi chain
(see text). The expected amplitude of the supercurrent is shown as
dotted lines in trace (a).}
 \label{envelop}
\end{figure}

The different histograms shown in Fig.\ref{oscillation}c  illustrate how the switching probability evolves within one fast period of the $I_{SW}$ sawtooth. The sharpest histogram is obtained in the middle of the linear sawtooth $i.e.$ when the supercurrent goes to zero (minimum of energy in the parabolic diagram shown in Fig.\ref{chainstates}a). For this point the state of the chain is quite stable. The presence of two steps in the histogram near the maximum or minimum of the sawtooth could be an indication that the system can switch between the states $\left|m\right\rangle$ and $\left|m+1\right\rangle$.  They reveal the crossing of energy levels between successive parabolic arcs of the energy diagram. The whole plot evolves slightly when the criterion for the definition of the switching current is set different from $50\%$ but the main features are preserved.
The observed behaviour at 26mK is characteristic for the zero temperature limit. We saw no change with increasing moderately the temperature. The trace remained very similar except for a small change in the vertical scale. For example at $326$ mK the amplitude of the fast oscillation was found to decrease by about $6\%$ and $10\%$ for the $h/2e$ and the $h/4e$ components respectively. We also observed very rare flux jumps which manifest themselves as discontinuities in the $I_{SW}$ $vs$ $\gamma$ curve. Further reduction of the oscillation amplitude was observed at higher temperature (up to $0.8 K$) together with some thermal smearing. \\

Practically we analyze the $\gamma -$dependence of the switching
current as the sum of 3 distinct contributions : a constant level
that can be assigned to the switching current of the shunt
junction, a fast oscillation due to the persistent current in the
large superconducting loop containing the large junction and an additional contribution reminiscent of the
switching current of the open chain. In Fig.\ref{envelop}a, we
have extracted the fast oscillating component  $I_{s}$ of the
measured switching current of sample A from the median line
$I_{med}$ obtained by joining the middle points of each branch of
the sawtooth in Fig.\ref{oscillation}a. The median line
(Fig.\ref{envelop}b) is reminiscent of the switching current of
the reference open rhombi chain which was measured separately
(Fig.\ref{refA}). The exact cause for this resemblance is not yet understood. From
the measurements we estimate the switching current of the shunt
junction at $1.43 \mu A$, which looks reasonable.

The fast oscillating component is shown in Fig.\ref{envelop}a. The
main features of this experimental result follow the theoretical
predictions summarized in Fig.\ref{chainstates}. Since by
changing the magnetic field we vary in the same time the
frustration and the phase, we obtain supercurrent oscillations
with a modulated amplitude. In Fig.\ref{envelop}a we have also
plotted the theoretical envelop $I_{env}$ (dotted lines) of the supercurrent
as calculated for the actual junction parameters in the classical
limit. This line is given by the maximum of the supercurrent $I_{S}(\gamma)$ and except for the two small windows visible near $f=\pm 1/2$ it is given by $I_{env}= i_{c} \frac{\pi}{N} \cos(f/2)$ (here $N=8$). Within the frustration window (see eq. (\ref{window}) and Fig.\ref{sawtooth}), $ I_{env}$ falls linearly to its minimum
value $i_{c} \frac{\pi}{2\sqrt{2}N}$ at $f=\pm 1/2$, as theoretically expected.

As it can be seen, the measured amplitude of the supercurrent
coincides fairly well with the classical value obtained from the
nominal critical current of the individual junctions. It appears
that the rate of quantum phase slip is too slow to achieve the
quantum superposition of classical states and form the macroscopic
$4e$ condensate. This is not surprising considering that the ratio
$E_{J}/E_{c}=20$ is significantly larger than the optimum values
calculated in \cite{Protopopov_04}. No rounding or exponential
weakening of the sawtooth-like supercurrent is observed. Rather we
do see the signature of the succession of classical states forming
the ground state illustrated in Fig. \ref{chainstates}. The same
is true for sample B.

The detailed field dependence of the fast oscillation contribution
can be very well understood from the classical ground state of the
phase biased rhombi chain. Fig.\ref{sawtooth} displays the experimental
switching current $I_{SW}$ together with the calculated
supercurrent near $f=1/2$ for sample B. This sample has the
largest ring area and therefore the largest number of fast
oscillations. We observe the emergence of the half period in a
frustration window $0.447<f<0.553$ as expected from eq.
(\ref{window}) for $N=8$ rhombi. Some additional secondary cusps,
presumably due to flux jumps are also observed in the experimental
trace.

\section{Quantum chains}
\begin{figure}[htbp]
\includegraphics[width=15cm]{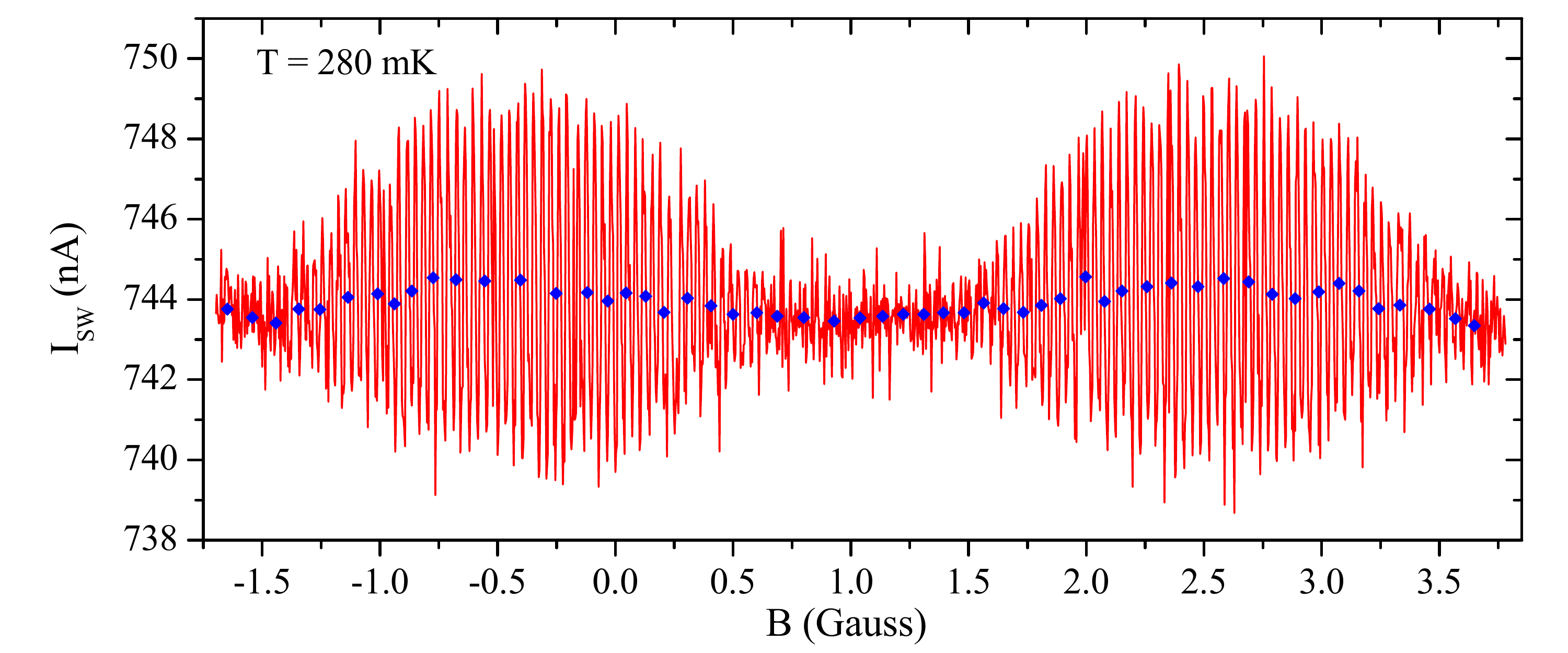}
\caption{(Color online) In red: the experimental plot of the switching current $vs$ external
magnetic field for the sample C at the temperature $T=280 mK$.
Blue points: the median component of the switching current.}
\label{DeuxPeriodes}
\end{figure}

In order to characterize the regime of quantum fluctuations,
experiments on rhombi chains with a ratio of $E_{J}/E_{C}\approx
2$ were performed. The measured histograms, unlike in the case of
the classical chains, do not split into steps. Such a behavior is
expected in the case where a significantly large gap opens in
between the classical states at the cross over point, and thus it
prevents the excitation of the system. In our case however, the
width of the histograms of $\approx 45 nA$ is much larger than the
amplitude of the switching current oscillations (see
Fig.\ref{DeuxPeriodes}). So even if there were some transitions
twards the first excited state (measurement induced or thermal excitations, noise), the splitting
of the histograms would not be visible.

Fig.\ref{DeuxPeriodes} shows the dependence of the measured
switching current as a function of the applied magnetic field. As
in the case of the classical chain, the signal can be seen as a
superposition of three components. The modulated oscillating
component characterizes the dependence of the supercurrent of the
chain as a function of both the frustration and the phase
difference $\gamma$. The oscillations are periodic with period
$h/2e$. As we approach the frustrated regime no oscillations of
the supercurrent are measured: in the region $f=1/2$ the
supercurrent of the chain is strongly suppressed and smaller than
the $\approx 1 nA$ noise of our experiment.

\begin{figure}[htbp]
\includegraphics[width=15cm]{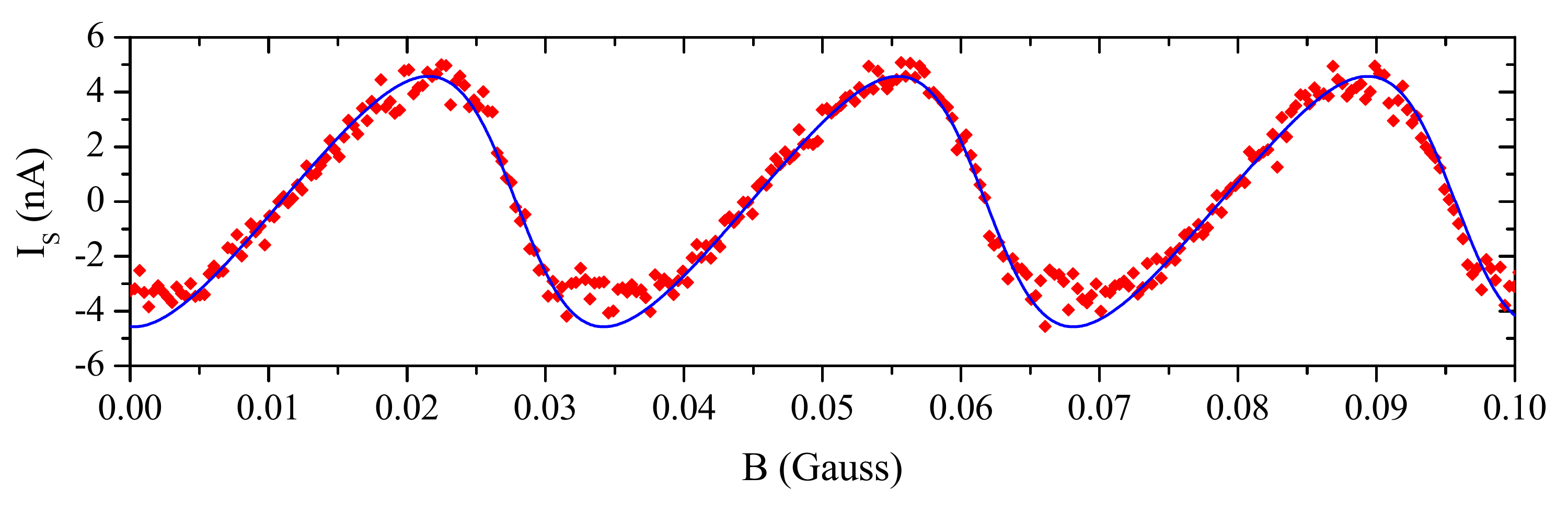}
\caption{(Color online) The experimental plot (red points) of the switching
current $vs$ external magnetic field in the zero frustration region
for the sample C at $T=280 mK$. The line (in blue) represents the
theoretical fit which gives an effective value for the Josephson
energy $E^{*}_{J} = 0.5 E_{J}$.}
\label{fEgalZero}
\end{figure}

\begin{figure}[htbp]
\includegraphics[width=12cm]{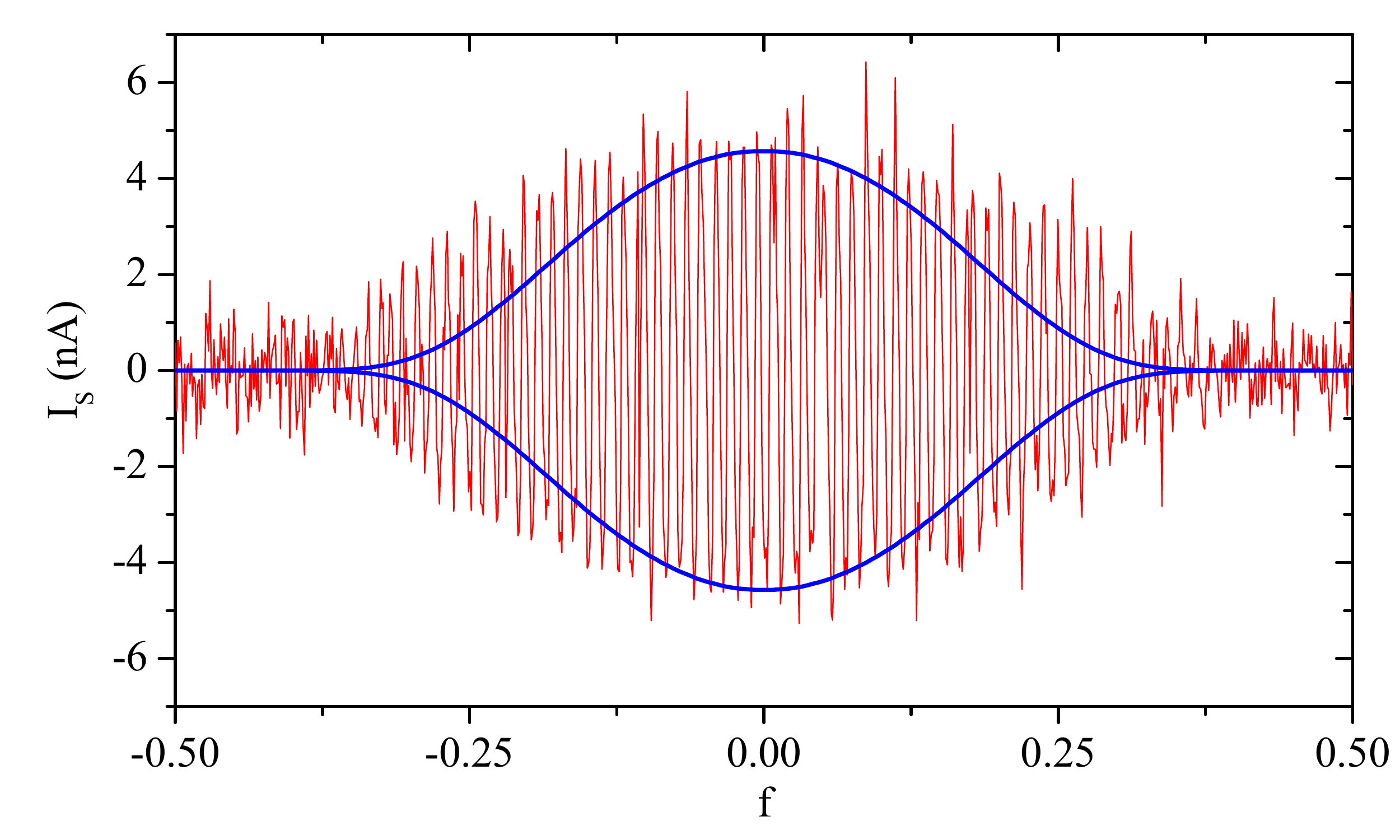}
\caption{(Color online) In red: the measured switching current oscillations as
a function of the frustration $f$. The blue line gives the theoretical
prediction for the amplitude of the switching current oscillations by
using an effective value $E^{*}_{J}=0.5 E_{J}$.}
\label{fDeZeroAUn}
\end{figure}

The median component $I_{med}$, shown in Fig.\ref{DeuxPeriodes},
as in the case of the classical chains, shows a periodic evolution
as a function of the frustration. We measure an amplitude of about
$1 nA$ for the $I_{med}$ oscillations. The exact cause of this
periodic behavior for chains where the phase difference is fixed,
was not yet understood. We did a detailed quantitative analysis of
the current phase relation at zero flux frustration. Fig.\ref{fEgalZero} shows the measured current phase relation in the non frustrated regime that can be perfectly fitted by the theory
described in section III, part B. The only fitting parameter is
the Josephson energy $E^{*}_{J}$ for which we find half of the
experimental determined one. We can imagine two possible sources
for this discrepancy. Firstly the experimental value for
$E^{*}_{J}$ has been deduced from the normal state resistance
measurement of the large Josephson junction (that is in parallel
to the rhombi chain) by supposing the ratio between the two
resistances to be the same than the one between the junction
areas. This assumption is not always valid in the case that
oxidation can occur differently for small junctions than for
larger ones. The second source of discrepancy could originate in
applicability of the theory described in section III. Formally the
description presented above relies on the assumption $E_J\gg E_C$.
On the other hand, even for $E_J\sim E_C$ the matrix $\upsilon$
for the single tunneling event is much smaller than $E_J$. This
means that we still can describe the system with the tight-binding
Hamiltonian (\ref{Ham}) but the precise value of $\upsilon$ can
deviate from the one given by equations (\ref{upsilon}, \ref{S_0},
\ref{A}).

To the best of our knowledge this result constitutes the first experimental confirmation of the model proposed by Matveev \textit{et al.} \cite{Matveev_02} for the current-phase relation in long Josephson junction chains. 

As we increase the applied magnetic field, the frustration inside
the rhombi modifies the value of the effective Josephson energy,
which becomes $E_{J}cos(\pi f)$. Using this value, we calculated
the evolution of the critical current as a function of the
frustration $f$. Fig.\ref{fDeZeroAUn} presents both the results
of the calculations and the measured values for the critical
current. We can see that the model gives a quantitative
description for the measured current amplitude dependence in the
non frustrated regime while it can only give a qualitative
description in the frustrated region.

\section{Conclusion}
In this paper we have studied the properties of one-dimensional
Josephson junction chains where the elementary cell is a rhombus
made of 4 small tunnel junctions. In the classical phase regime, the current-phase relation shows the
characteristic sawtooth-like variation. Its periodicity
corresponds to the ordinary superconducting flux quantum $h/2e$
when the rhombi chain is non frustrated and it turns to half the
flux quantum $h/4e$ at maximum frustration. For large $E_{J}/E_{C}$
ratio the observed current-phase relation can be well understood
from the classical ground state of the chain. The latter consists of a
sequence of successive parabolas differing by the entrance of
phase slips into the chain. Experiments on rhombi chains in the
quantum regime ($E_{J}/E_{C}\approx 2$) show a significant
reduction and rounding of the current-phase relation in the
non frustrated region and a complete suppression of the supercurrent at maximal frustration. In the non frustrated regime we were able to apply for the first time the model proposed by Matveev \textit{et
al.}\cite{Matveev_02} in order to successfully fit the measured current phase-relation for an eight rhombi quantum chain.

\section{Acknowledgment}
The authors are grateful to B. Dou\c cot, M. Feigelman and L. B.
Ioffe for many inspiring discussions. We are indebted to Th.
Crozes for his help in sample design and fabrication. The samples
were realized in the Nanofab-CNRS platform. We also acknowledge
the INTAS program "Quantum coherence in superconducting
nanocircuits", Ref. 05-100008-7923  for financial support.

\section{Appendix : characterization of open chains}
In this appendix we consider the transport properties of current biased
Josephson chains connected to external reservoirs. Since the
chains are open, the phase condition given in eq. (\ref{phasechain})
does not hold. In this configuration, the switching current of the circuit $I_{SW}$ coresponds to the maximum supercurrent through the chain and it strongly depends on the frustration. 
We have measured the current-voltage characteristics of chains based on three different elementary cells: a single junction, a SQUID or a rhombus, with lengths varying from $N=1$ to $N=64$. The range of junction parameters is the same as in the main part of this paper. Our general observations are the following :\\

The current voltage characteristic of chains with large Josephson coupling energy
($E_{J}/E_ {C} \gg 10$) is similar to that of a single cell with a multiplicative factor $N$ in voltage. The I-V
characteristics are strongly hysteretic and, for small $N$, the
switching current is close to the Ambegaokar-Baratoff value.  In
rhombi chains, the switching current is periodic with respect to
the frustration, in particular it drops by a factor 2 at
frustration $1/2 $ as expected from eq. (\ref{icrhombus}). The SQUID
chain exhibits the usual sinusoidal dependence with full
cancellation of switching current at frustration $1/2$. It behaves
as a chain of single junctions which Josephson energy is tuned by
the external magnetic flux. The observation of a fully developed
critical current indicates that the chains can be seen as a series
of independent cells which remain in a metastable
state of energy much higher than the ground state energy shown in
Fig.\ref{chainstates}a for the closed chain. This fact is not
surprising since the energy barrier for phase
jumps is very high in these strong chains.\\

\begin{figure}[htbp]
\centering
\includegraphics[width=8cm]{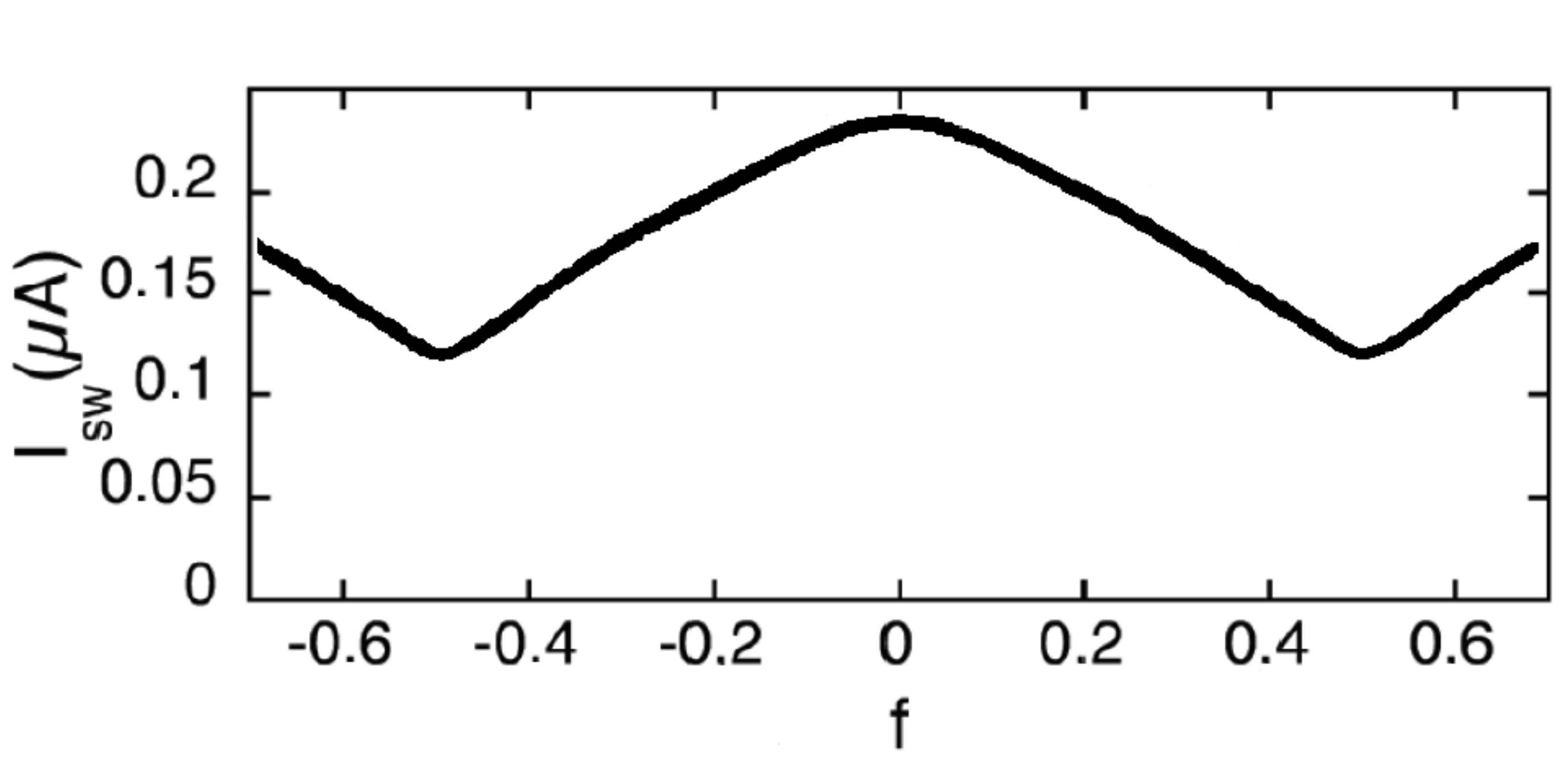}
\caption{Switching current of a $N=8$ open
rhombi chain $vs$ frustration. The $E_{J}/E_{C}$ ratio is 27
and the Ambegaokar-Baratoff critical current is $0.74\mu A$.}
\label{refA}
\end{figure}

Increasing the length or decreasing the Josephson coupling results
in a dramatic reduction of the switching current. For example
Fig.\ref {refA} shows the switching current measured in a rhombi
chain with $N=8$, made with identical fabrication parameters and on
the same chip as sample A (see Table \ref{sampleswitch}). The
zero field switching current is about $1/3$ of the
Ambegaokar-Baratoff value although the ratio $E_{J}/E_{C}$ is
large. The I-V characteristic for this class of samples is
hysteretic except near full frustration.\\

\begin{figure}[htbp]
\centering
\includegraphics[width=8cm]{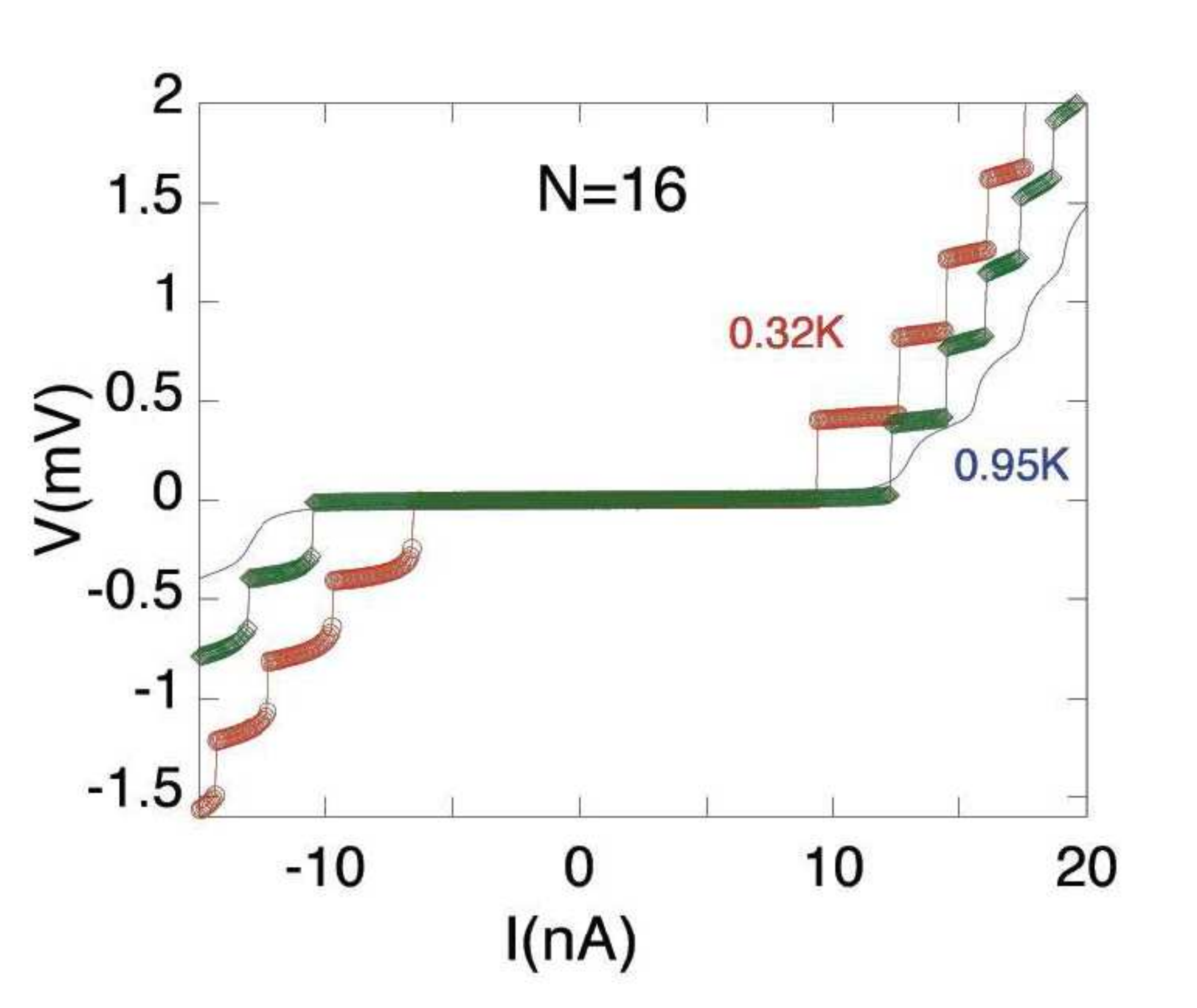}
\caption{(Color online) Current voltage characteristics of a weak coupling N=16
Josephson chain at temperatures $T=0.32$, $0.8$, and $0.95K$. The elementary voltage jump coresponds to twice the value of the superconducting gap $ \Delta=200 \mu eV $. The ratio $E_{J}/E_{C}\approx 6$ and the Ambegaokar-Baratoff critical current of a single junction $i_{c}\approx 100 nA$.}

\label{JJ16}
\end{figure}

A distinct behavior is found in weaker junctions ($E_{J}/E_{C}
\leq 10 $), when the rate of thermal and quantum phase slips is
significant at the time scale of an experiment. On the same chip we fabricated a set of chains where the elementary cell is formed by a single Josephson junction of area $0.15 \times 0.3 \mu m^{2}$. The chains contained respectively 1, 4, 16 and 64 jonctions. The tunnel resistances per individual junction
were found almost identical $r_{n}=3 \pm 0.2 k\Omega$ which is an
indication of good homogeneity of the array. We observed step-like
characteristics with voltage jumps equal to the superconducting
gap $2\Delta$. Each jump corresponds to the switching of one
junction, see Fig.\ref{JJ16}.  We identify the switching current
$I_{SW}$ at the first jump, $i.e.$ when the weakest junction runs
into a voltage state. For $N=16$, $I_ {SW}$ is about 10 times
smaller than the expected single junction
critical current.\\

We found the following characteristics as a function of the chain length :\\
- The switching current reduces as the chain length $N$ increases : $23, 10$ and  $1.2$ nA respectively for $N=4, 16$ and $64 $. We also observe that $I_{SW}$ increases with increasing temperature, indicating that the thermal fluctuations restore the phase coherence of the chain by suppressing quantum processes.\\
- The hysteresis of I-V curves is suppressed in long chains, giving rise to a regular staircase shape.  \\
- The Josephson branch becomes more and more dissipative as the
chain length increases. The measured zero bias resistances are
respectively $40 $, $380$ and $50 k\Omega$ for $N=4$, $16$ and
$64$. Further reduction of $E_{J}$ leads to the total suppression
of the Josephson coupling in the chain.\\
 The very regular sequence of steps cannot be due to sample
inhomogeneities. We believe  that the local environment of the
different junctions inside the chain plays a significant role. Preliminary
experiments where a Josephson chain was shunted by a on-chip
interdigit capacitance ($1 pF$) did not reveal any significant change.\\
Up to now we have no quantitative understanding of these observations on open chains.


\begin{thebibliography}{99}
\bibitem{gamma} $\Phi_{c}$ is the magnetic field times the average area defined by the internal and external contours of the rhombi chain. To be specific, our phase $\gamma$ coincides  with the term $\tilde{\gamma}$ defined in Ref\cite{Protopopov_04}.
\bibitem{homogeneity} We noticed in our previous designs a few $\%$ difference in the area of neighbour junctions leading to a small systematic inhomogeneity of critical currents. Here we made a new design which insures that any possible drift of the e-beam induced identical deformation on the shape of each junction. Dispersion of junctions sizes was finally not detectable.
\end{thebibliography}
\end{document}